\documentclass[10pt]{iopart}
\usepackage{iopams}
\usepackage{amsmath,amssymb,color}
\usepackage{graphicx}
\usepackage{epsfig}
\usepackage{color}

\newcommand{\ie}{\emph{i.e.}, }
\newcommand{\reff}[1]{(\ref{#1})}
\newcommand{\ereff}[1]{Eq.(\ref{#1})}
\newcommand{\erefs}[1]{Eqs.(\ref{#1})}
\newcommand{\figref}[1]{Figure \ref{#1}}
\newcommand{\Ex}{E_{n}^{x}}
\newcommand{\Ey}{E_{n}^{y}}
\newcommand{\modu}[1]{\mid\!\!{#1}\!\!\mid}
\newcommand{\p}{\partial}
\newcommand{\fb}{\bar{f}}
\newcommand{\fqss}{\bar{f}_{QSS}}
\newcommand{\fo}{f_{0}}
\newcommand{\gs}{\,\,{\scriptscriptstyle{\gtrsim}}\,\,}
\newcommand{\ls}{\,\,{\scriptscriptstyle{\lesssim}}\,\,}
\newcommand{\omzn}{\omega_{0n}}

\newcommand{\emphr}[1]{{#1}}

\begin{document}

\title[Beam-Plasma Instability and Fast Particles: the Lynden-Bell Approach]{Beam-Plasma Instability and Fast Particles: \\the Lynden-Bell Approach}

\author{Nakia Carlevaro$^{1}$, 
Duccio Fanelli$^{2}$,
Xavier Garbet$^{3}$,
Philippe Ghendrih$^{3}$, 
Giovanni Montani$^{4,1}$ and 
Marco Pettini$^{5}$}

\address{ \vspace{0.5cm} 
$^{1}$ Department of Physics, ``Sapienza'' University of Rome (Italy), \\P.le Aldo Moro, 5 (00185) Roma, Italy;\\
$^{2}$ Department of Physics and Astronomy, University of Florence and INFN,\\ Via Sansone, 1 (50019) Sesto Fiorentino, Italy;\\
$^{3}$ CEA Cadarache - IRFM, (13108) Saint-Paul-les-Durance, France;\\
$^{4}$ ENEA - C.R. Frascati UTFUS-MAG, Via E. Fermi, 45 (00044) Frascati, Italy;\\
$^{5}$ Centre de Physique Th\'eorique, UMR 7332, Aix-Marseille University,\\ Campus de Luminy, Case 907
(13288) Marseille cedex 9, France.}

\begin{abstract} The beam-plasma instability, \ie the response of the plasma bulk to the injection of supra thermal charged-particle beams, results to be appropriately characterized by a long-range interaction system. This physical system hosts a number of very interesting phenomena and, in particular, the emergence of long-lived quasi-stationary states. We characterize the self-consistent distribution functions of such out-of-equilibrium states by means of the Lynden-Bell's theory. The prediction of this theory, based on the statistical mechanics of the Vlasov equation, are checked against the outcomes of numerical simulations of the discrete system. Moreover, a phenomenological study of the effective resonance band for the system response is also addressed. A threshold value has been found in the initial spread of beam-particle momenta. This threshold allows to discriminate between the resonant and non-resonant regimes. The analysis of the thermalization of a \emphr{few percents of the beam population characterized by large initial momenta (with respect to the main part of the beam itself)} is also performed and it confirms and deepens the understanding of the physical meaning of the mentioned threshold.
\end{abstract}


\section{General background}
\label{intro}

Long-range interactions are characterized by two-body interaction potentials which are inversely proportional to a power of the inter-particle distance which is smaller than the number of spatial dimension (otherwise, they are said to be short-range interactions) \cite{A09}. In this case, it can be shown that the energy per particle diverges for large sizes and the contribution due to the boundary surfaces of the system cannot be neglected with respect to the bulk. This implies that the evolution of an element of the treated model is affected by all the other ones, instead of by its neighbors only, as is the case of short-range interactions. Several recent physical experiments involving long range interactions call for the search of novel theoretical tools to describe the physics these systems.

One of the most important features (beyond the ensemble inequivalence, the presence of negative specific heat and broken ergodicity \cite{A09}) emerging out of long-range interactions is the presence of the so-called quasi-stationary states (QSSs). Such regimes correspond to apparently stable, long-living out-of-equilibrium states whose life-time diverges with the number of constitutive elements of the system. These regimes are followed by a slow relaxation to thermal equilibrium \cite{B04,PLR04}. This emergent property is encountered in several physical contexts where long-range forces are present, \emph{e.g.}, Free-Electron Laser, Travelling Wave Tubes and, in general, in plasma-wave models. Moreover, the conceptual relevance of these long-living states stems from the argument that these are the only experimentally accessible regimes. \emphr{Other examples of states living for large time out of the equilibrium can be found in the literature. In particular, steady holes were found for ion acoustic turbulence \cite{SK96} and for the bump-on tail instability \cite{LI12}. These models are self-consistent, \ie the electric field is dynamically coupled to the particle response, but the holes that develop are (weakly) dissipative structures (of collisions matter). This is particularly true for \cite{LI12}, where long-lived phase space holes were prominent when the drag term in the collision operator is dominant. At variance, the QSS treated in this paper can be obtained for a class of collisionless dynamical systems, and therefore they are not dissipative structures.} Even though several attempts have been proposed for a systematic theoretical description of QSSs, a complete and satisfactory theory is still lacking. Among the approaches so far proposed, two of them deserve to be mentioned: a maximum entropy principle based on the Lynden-Bell's theory \cite{LB67}, and the approach introduced in \cite{LPT08} which moves from the observation that the system develops a  core-halo structure, induced by  parametric resonance. We will hereafter focus on the former one.

In this paper, we focus on a specific system involving long-range interactions: the beam-plasma instability model \emphr{\cite{OWM71,EEb}}. This model describes the injection of a supra thermal electron beam into a thermalized plasma. Under particular conditions, the plasma response is non-resonant and it can be treated as a bulk with respect to the propagation of the beam. Nonetheless, the system gets unstable by generating Langmuir modes, and charged particles are effectively long-range-coupled one to the other via these electrostatic waves. After an initial phase of growing amplitude of the Langmuir modes, the particles get resonantly trapped and the system attains an oscillatory regime of the wave intensity. In this respect, by performing a phenomenological analysis of the resonance band through the study of the beam initial conditions, it is shown that a threshold value in the momentum space exists for which the system responds to the beam injection by clustering the beam particles in rotating clumps. In other words, a rather sharp transition between the resonant and non-resonant regimes is observed. Moreover, the existence of this threshold is consistent with the analysis concerning the thermalization of a small (few percent) population of beam particles having large initial momenta with respect to the main part of the beam itself.

As discussed above, the resonant system achieves an intensity-oscillatory regime and we will show that this dynamical phase corresponds to a QSS. In this respect, a relevant contribution of the present work is the achievement of a consistent theoretical picture able to properly characterize such peculiar intermediate states. In particular, by analyzing the continuum limit of the discrete beam-plasma system, we show that the macroscopic observables associated to the QSSs are accurately predicted by a maximum entropy principle inspired to the Lynden-Bell violent relaxation theory \cite{LB67}, \ie an analytical treatment based on the equilibrium of the associated Vlasov equation. Such an approach relies on the observation that, in the continuum limit (corresponding to an infinite number of particles), the discrete set of equations converges toward the Vlasov equation, which describes the evolution of the particle distribution function. Within this scenario, the Lynden-Bell theory surmises that the QSSs correspond to the statistical equilibria of the continuous Vlasov model, and the validity of this \emph{ansatz} is verified in this paper. In fact, it is only after testing the consistency of this conjecture, that the continuum model can be legitimately used to predict (with a good accuracy) the relevant observables of the QSSs. In particular, it provides an analytical form of the distribution function of the out-of-equilibrium transient states. Actually, the increasing interest for the role played by fast particles populations in affecting the confinement properties of a burning thermonuclear plasma, calls attention to the construction of integrated algorithms for the comparison between theory and experiment. In this perspective, relevant contribution can be given by predictions based on numerical simulations and by fundamental theories addressing simplified yet physically non trivial models to characterize the long time-scale non-linear dynamics of burning plasmas that will be produced in the forthcoming tokamak experiments. \emphr{In fact, one potential limitation in fusion devices comes from magneto-hydrodynamical instabilities driven by energetic particles (MHD-EP), \emph{e.g.}, alpha particles born from reactions fusion. In principle, the proper kinetic description of the MHD-EP instabilities requires to solve the three dimensional Maxwell equations coupled to a Vlasov equation for fast particles, in a 6-dimensional phase space. The particle trajectories in the unperturbed magnetic configuration of a tokamak can be described by a set of action/angle variables, and it has been shown in \cite{BB99} that if only one magneto-hydrodynamical instability is involved, a proper change of conjugate variables allows a reduction to a 1-dimensional problem (2 dimensions in phase space), which belongs to the class of bump-on tail instabilities. Hence, we stress that the model presented in this paper is relevant to MHD-EP instabilities in tokamaks in that framework.}

This paper is organized as follows. In Section \ref{BPI-equations}, the basic equations describing the discrete beam-plasma system are derived \cite{TMM94}. In Section \ref{Simulations}, the outcomes of simulations performed with uniform initial conditions concerning the particle and wave-intensity behaviors are reported pointing out the existence of the QSSs and of the resonance threshold. In Section \ref{Vlasov-LyndenBell}, the continuum limit is addressed and the corresponding Vlasov equation is numerically integrated using the Lynden-Bell approach. The distribution function (which is derived to characterize the QSSs) is calculated and plotted together with the wave intensity. In Section \ref{Test}, the Lynden-Bell model is tested by comparing the theoretical continuum results to numerical simulations and a satisfactory agreement is found. The concluding remarks follow.

\section{Self-consistent Hamiltonian for the beam-plasma instability}
\label{BPI-equations}

The system we analyze in this paper involves self-consistent long-range interactions by considering non-interacting charged particles subjected to a time-dependent potential which, in turn, feels the particle influence. This generates the long-range self-consistent coupling and each particle interacts with the others through the field and not directly, at variance with the fully self-consistent $N$-body dynamics.

In particular, we address the beam-plasma instability problem. The model consists of a beam of charged particles moving in a background neutral plasma. When a supra thermal electron beam is injected into a thermalized plasma, the system reacts by forming electrostatic plasma waves, \ie a series of travelling sinusoidal potentials which are called Langmuir waves \cite{EEb,AEE98,O65}. Such electrostatic potentials oscillate at a pulsation close to the plasma frequency (in view of the higher ion inertia) defined by $\omega_p^{2}=e^{2}n_e/\epsilon_0 m_e$, where $n_e$ is the electron density of the plasma, $e$ the electron charge (assuming hydrogen like ions), $m_e$ the electron mass and $\epsilon_0$ the vacuum permittivity. \emphr{We underline that the model is discussed in the reference frame comoving with the injected beam \cite{OWM71}.}

\subsection{Single-wave model}

The specific approach to the beam-plasma instability we are going to generalize is the single-wave model proposed in \cite{OWM71} and \cite{M72,S63} (the system has been also extensively studied in \cite{FE98,FE00,F01}). In this picture, the dynamics described above is analyzed by considering that the most unstable mode leads the dynamics and hence by neglecting the other modes. This assumption is well motivated only during the linear evolution and our generalization consists of the description of the non-linear dynamics when a second dominant mode is introduced. In fact, it is expected that in the forthcoming laboratory plasma experiments, like the ITER tokamak, tens of modes will be excited depending on the plasma parameters
In \cite{OWM71}, it was shown how the wave amplitude grows until the particles of the beam are trapped by the wave and then, as it saturates, it starts  oscillating because the beam resonantly responds to the potential. The model considered applies to the case of one dimensional, non-relativistic, electrostatic system where the response of the plasma and of the beam are treated separately. In particular, the phase velocity of the wave is assumed to be larger than the velocity of the background particles, thus the plasma responds non-resonantly and trapping effects of plasma particles (not belonging to the beam) can be reliably neglected. \emphr{Under this hypotheses, the plasma represents a non-resonant bulk responsible for the presence of the electrostatic modes and it is described by the linear plasma dielectric function \cite{OWM71,TMM94,EEb}, while the dynamics is specified for a beam composed by $N$ particles (labeled by the index $j$).} The system is described by the following set of equations:
\begin{subequations}\label{SWM}
\begin{align}
&\frac{d x_j}{dt}=\frac{p_j}{m}\;,\phantom{\sum_{j=1}^{N}}\\
&\frac{d p_j}{dt}=-e[E_{n}\;e^{ik_{n}x_j}+E^{*}_{n}\;e^{-ik_{n} x_j}]\;,\\
&\frac{d E_{n}}{dt}=-i\omzn E_{n}+\frac{4\pi e}{k_n L\epsilon^{\scriptscriptstyle\prime}}\;\sum_{j=1}^{N} e^{-ik_{n}x_j}\;,\label{SWM3}
\end{align}
\end{subequations}
where $\omzn$ denotes the electric field pulsation that characterizes the zeros of the dielectric function $\epsilon(k_n,\omega_n)$, \ie $\epsilon(k_n,\omzn)=0$, and $x_{j}$ and $p_{j}$ are the beam-particle positions and momenta, respectively. $E_{n}$ represents the electrostatic field, where the index $n$ labels the most unstable mode by the relation $k_{n}$ expressed in units of $2\pi n/L$, here $L$ defines the periodicity length of the system (it is worth recalling that the original analysis takes into account only one spatial Fourier component, \ie $k_n$ is fixed). \emphr{In \ereff{SWM3}, $\epsilon^{\scriptscriptstyle\prime}$ indicates the derivative of the dielectric function with respect to the pulsation, \ie $\epsilon^{\scriptscriptstyle\prime}=\p\epsilon/\p\omega|_{\omega=\omzn}$. In fact, in the case of a weak beam the electrostatic response results mainly due to the zeros of $\epsilon$ and one can safely expand such quantity near $\omzn$ at first order. Moreover, the dispersion relation governing the electric field perturbations in a beam-plasma system can be written as a condition on the dielectric function of the beam, thus setting the morphology of the treated Langmuir wave (for the details of this specific analysis, see \cite{OM68}).
}

\erefs{SWM} form a closed dynamical system governing the (single) wave-beam interaction and it can be formally derived within an Hamiltonian framework. Such a characterization was first addressed in \cite{MK78} and then formally established in \cite{TMM94}. In particular, the self-consistent Hamiltonian formulation of the equations of motion was obtained from the Vlasov-Poisson equation (a direct derivation was instead proposed in \cite{EEb,AEE98}). In so doing the system can be modeled by a Hamiltonian describing the self-consistent dynamics of waves and beam particles, and the waves appear as harmonic oscillators. Moreover, non-resonant bulk particles have a trivial oscillatory motion and form a dielectric medium supporting the wave propagation. Using the restriction of the single space dimension introduced above, the Hamiltonian for the system \reff{SWM} writes \cite{TMM94}
\begin{align}\label{SWM_H}
H=\frac{\omzn L\epsilon^{\scriptscriptstyle\prime}}{4\pi}\modu{E_{n}}^{2}
+\sum_{j=1}^N \Big[\frac{p_j^2}{2m}-\frac{ie}{k_{n}}
(E_{n}\;e^{ik_{n}x_j}-E^{*}_{n}\;e^{-ik_{n} x_j})\Big]\;.
\end{align}

As already discussed, the whole dynamical single-wave model predicts a linear exponential instability and a later oscillating saturation for the amplitude of the electric field. A numerical analysis \cite{OWM71,TMM94} performed for a mono-kinetic beam indicates that the particles initially trapped by the wave behave like a large ``macro-particle'', that evolves coherently. This scenario allowed the introduction of a simplified Hamiltonian model \cite{TMM94}, with only four degrees of freedom, \ie the wave, the macro-particle and the two boundaries delimiting the portion of space occupied by the so-called ``chaotic sea''. The model has been revisited in \cite{AF06}, where a self consistent characterization of also the parameters involved was provided.

\subsection{Multi-wave model}
In the following, we address the generalization of the single-wave model scenario described above by still treating the Langmuir waves in the non-resonant plasma bulk as dynamical objects, \ie harmonic oscillators. In this sense, we use the transformations introduced in \cite{TMM94,A05} for rescaled and dimensionless variables based on the fundamental frequency $\omega_b^3=4\pi e^{2}N/mL\epsilon^{\scriptscriptstyle\prime}$, describing the dynamics in the presence of several wave modes.

\emphr{In particular, the evolution of a beam (composed by $N$ particles) interacting with $R$ electrostatic modes (labeled by the index $n$)} is found to be \cite{TMM94,FLA06} self-consistently governed by the set of the following equations (for the sake of simplicity, we drop the rescaled-variable notation)
\begin{subequations}\label{GWM}
\begin{align}
&\frac{d x_j}{dt}=\frac{p_j}{m}\;,\phantom{\sum_{j=1}^{N}}\\
&\frac{d p_j}{dt}=-\sum_{n=1}^{R}\,F_{n}[E_{n}\;e^{ik_{n}x_j}+E^{*}_{n}\;e^{-ik_{n} x_j}]\;,\\
&\frac{d E_{n}}{dt}=-i\omzn E_{n}+\frac{F_n}{N}\,\sum_{j=1}^{N} e^{-ik_{n}x_j}\;,
\end{align}
\end{subequations}
where $k_n$ denotes the unstable modes, while, since the model faces the electrostatic waves as harmonic oscillators, the corresponding coupling is described by $F_n$ defined in \cite{EEb} proportional to constant wave susceptibilities as $F_n=[\omega_p/2N]\sqrt{2mR/\epsilon^{\scriptscriptstyle\prime}\;}$. The pre-factor $1/N$ in the potential term prevents the occurrence of divergences that would appear in the plasma at large $N$, this pre-factor also keeps constant the plasma frequency for $N\to\infty$.  The above equations can be derived from the Hamiltonian \cite{EEb}
\begin{align}\label{GWM_H}
H=\sum_{j=1}^N\frac{p_j^2}{2m}+\sum_{n=1}^{R}\omzn \modu{E_{n}}^{2}
-i\sum_{j=1}^N \Big[\sum_{n=1}^{R}\frac{F_{n}}{k_{n}}
(E_{n}\;e^{ik_{n}x_j}-E^{*}_{n}\;e^{-ik_{n} x_j})\Big]\;.
\end{align}
In this scheme, the interaction of the Langmuir waves with  the electrons constituting the beam is characterized in the framework of a self-consistent Hamiltonian picture which is formally equivalent to the Free-Electron Laser dynamics \cite{CNSNS}.

\section{Numerical simulations}
\label{Simulations}

As previously discussed, the model we are considering describes a supra thermal electron beam injected into a thermalized plasma that destabilizes the electrostatic modes at the plasma frequency. Such a model is derived in the approximation of one-dimensional beam (the transverse components are neglected) and planar waves. \emphr{For the numerical analysis, we consider here only two electrostatic modes (\ie $R=2$). It important to stress that, in general, the ratio between the coupling constants $F_n$ depends on the details of the dielectric function and the morphology of the Langmuir waves. However, in this work, we analyze these quantities essentially to investigate the system response in correspondence to a different hierarchy in such parameters. Nonetheless it is possible to recognize that we consider values for $F_1$ and $F_2$ that can be obtained provided the density ratio between the beam and bulk to reach few percents, almost compatible with some configurations observed in experimental devices. The coupling constant order of magnitude could be refined by more detailed analyses using smaller values of the density ratio for a better match to most experimental conditions.} 

Explicitly writing the components of the generalized electric field in the form $E_n=\Ex+i\Ey$, the equations of motion \reff{GWM} read (for $m=1$)
\begin{subequations}\label{EqMotionSim}
\begin{align}
&\frac{d x_j}{dt}=p_j\;,\phantom{\sum_{j=1}^{N}}\\
&\frac{d p_j}{dt}=-\sum_{n=1,2}\,2F_{n}[\Ex\cos(k_n x_j)-\Ey\sin(k_n x_j)]\;,\\
\label{eq555}
&\frac{d E_{n}}{dt}=\frac{F_n}{N}\,\sum_{j=1}^{N} e^{-ik_{n}x_j}\;,
\end{align}
\end{subequations}
where we have neglected the wave contribution with respect to the particles motion, \ie $\omega_n\ll1$.
In this scheme, the canonically conjugated variables are $(x_i,\,p_j)$ and $(\sqrt{N}\,E_n,\,\sqrt{N}\,E_n^{*})$, moreover, the ``energy'' $H$ of \ereff{GWM_H} and the total ``momentum'' $P=\sum_j p_j +N\sum_n\modu{E_n}^{2}$ are conserved quantities.

\begin{figure}[!ht]
\centering
\includegraphics[width=0.45\hsize]{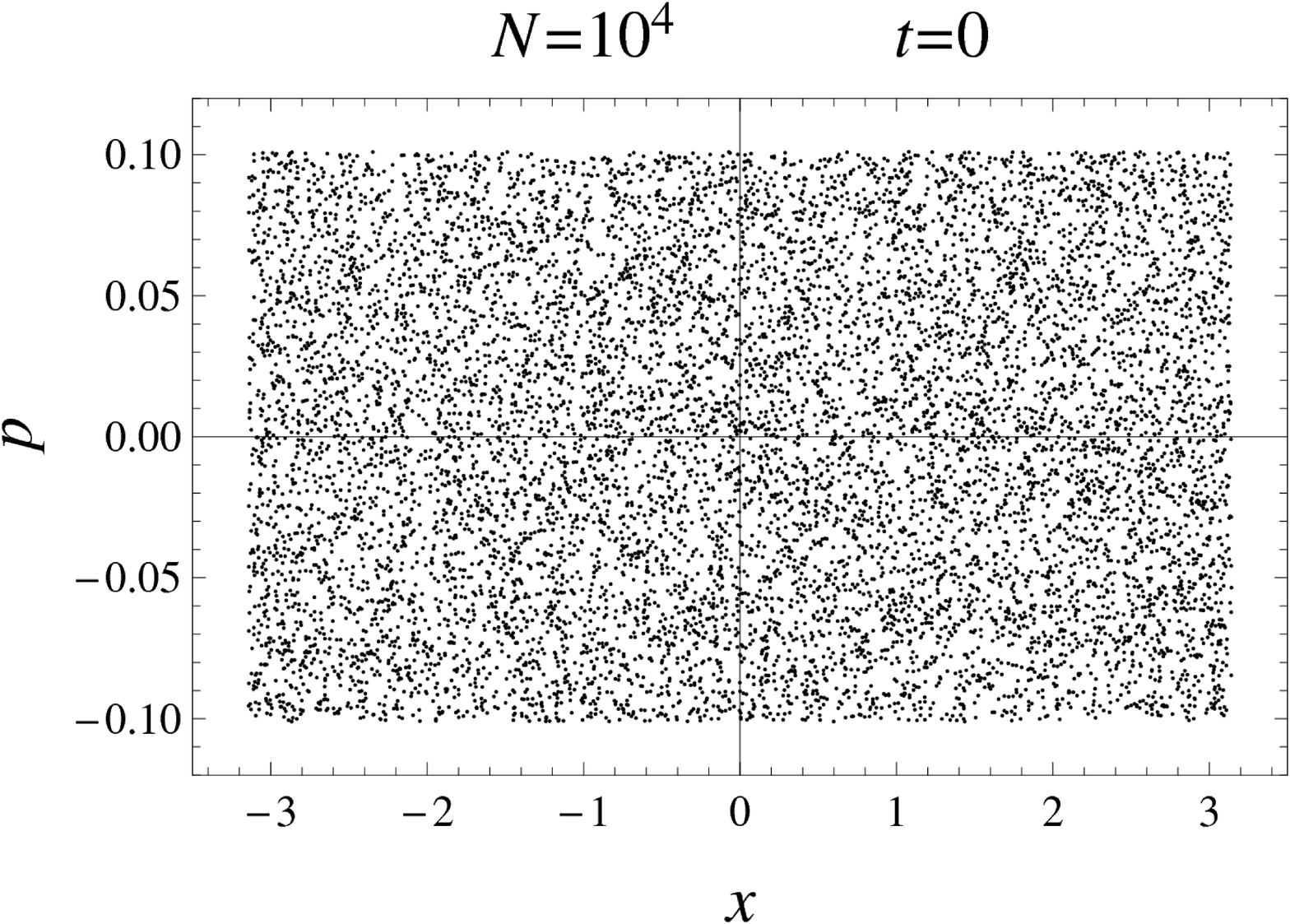}
\caption{Initial conditions in the $\mu$-phase space.}
\label{WaterBag0}
\end{figure}

In what follows, we will focus on the analysis of the electrostatic-wave intensity of the first and second harmonics, \ie $n=1,2$ ($k_n=1,2$, in units $2\pi/L$), defined by  $I_n=|E_n|^{2}=[(\Ex)^2+(\Ey)^2]$. The numerical simulations of the system \reff{EqMotionSim} are performed by considering as initial conditions (we recall that we are working in the frame moving with the injected beam) a random distribution of particles in the phase-space region defined by the constraint (arbitrary units)
\begin{align}\label{initialcond}
-\pi\leqslant x_j \leqslant\pi\;,\qquad
-p_0\leqslant p_j \leqslant p_0\;.
\end{align}
This choice corresponds to homogeneous initial conditions, since the bunching parameters $b_n=\sum_{j} e^{-ik_{n}x_j}/N$ (the Fourier coefficients of the single-particle distribution) are initially vanishing, \ie $\sin(\Delta x)/\Delta x=0$. Furthermore, the initial ``total energy'' is defined by $H/N=p_0^{2}/6$. In the following, we consider the values $N=10^4$ and $p_0=0.1$ (which corresponds to $H/N=0.0017$), as represented in \figref{WaterBag0} where the beginning of the dynamical evolution  is taken at $t=0$. Notice that, in what follows, the plots in the $(p,x)$ plane are snapshots of the particle distribution in the $\mu$-phase space, that is, each point represents the position $x$ and momentum $p$ of a particle at some given time. These figures have nothing to do (in spite of some resemblance) with Poincar\'e sections. 

\begin{figure}[!ht]
\centering
\includegraphics[width=0.44\hsize]{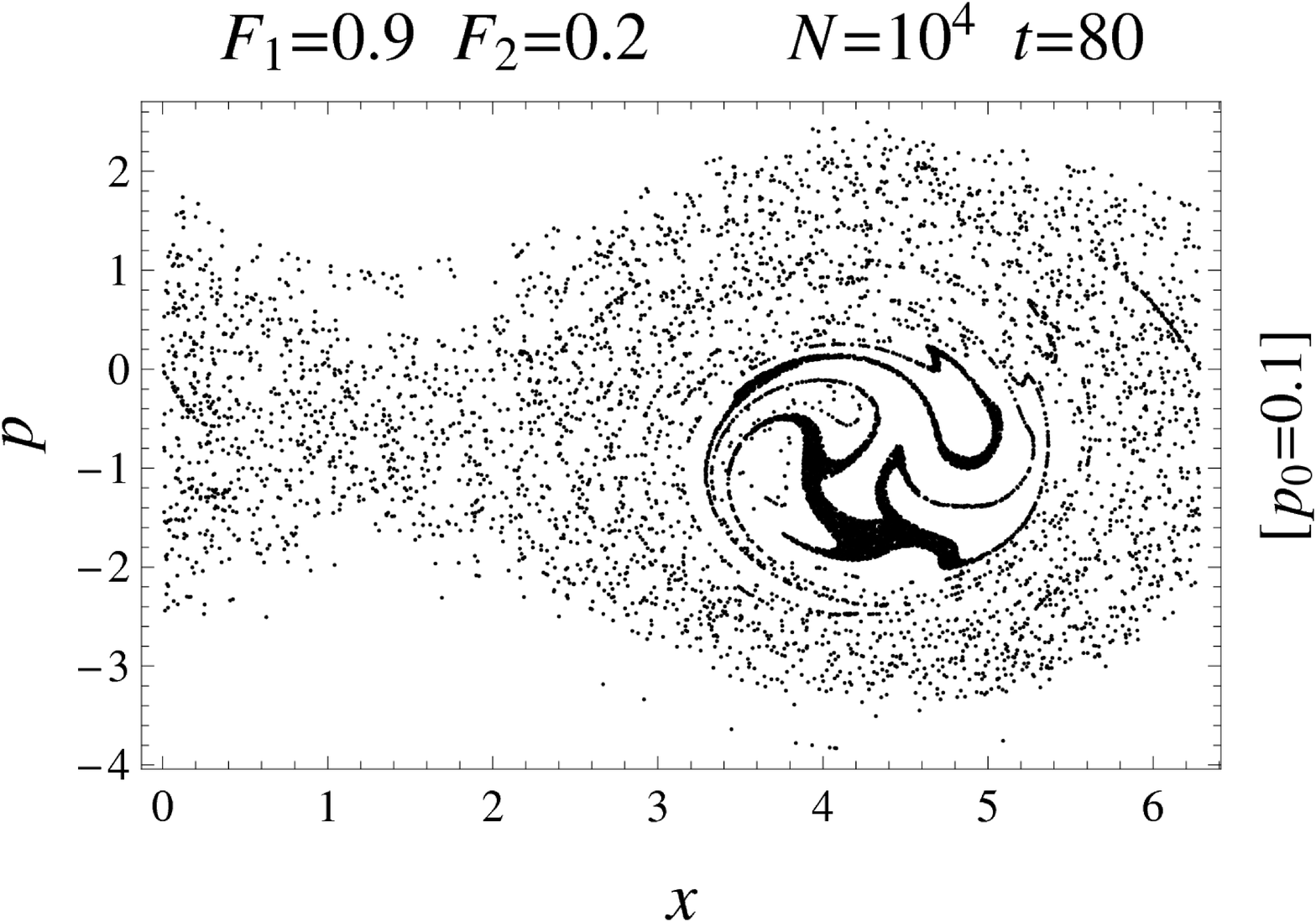}
\includegraphics[width=0.45\hsize]{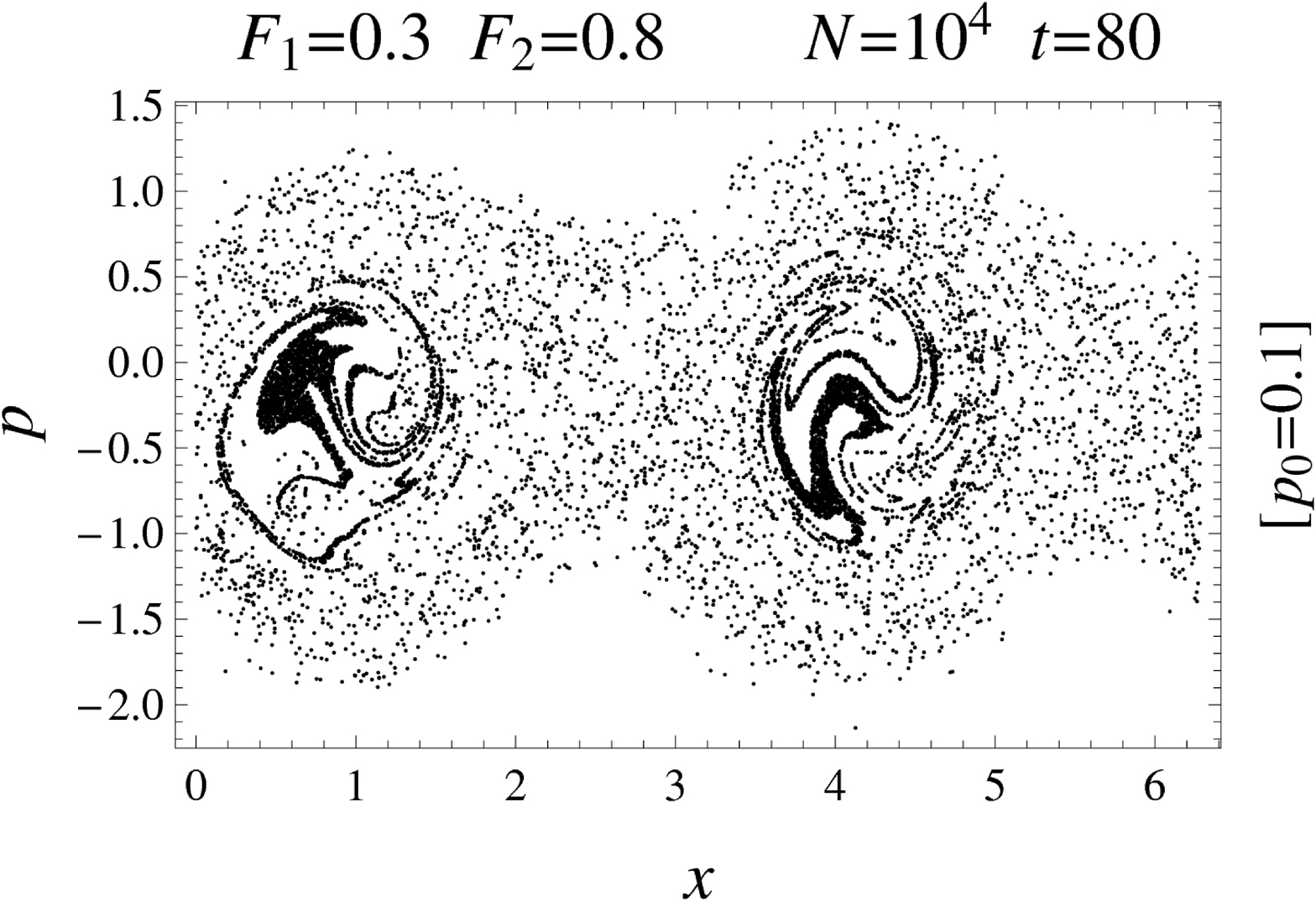}
\includegraphics[width=0.4\hsize]{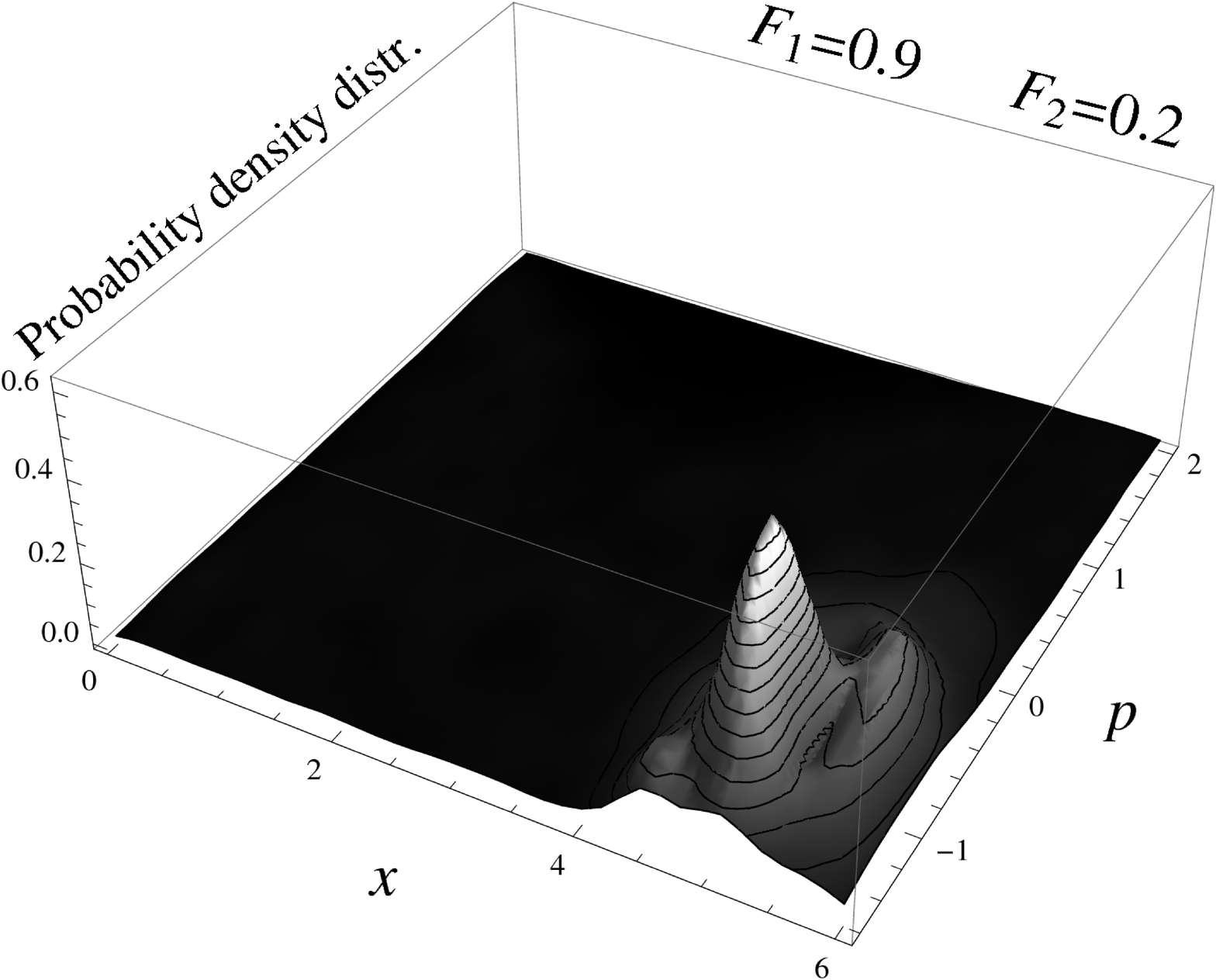}\quad
\includegraphics[width=0.4\hsize]{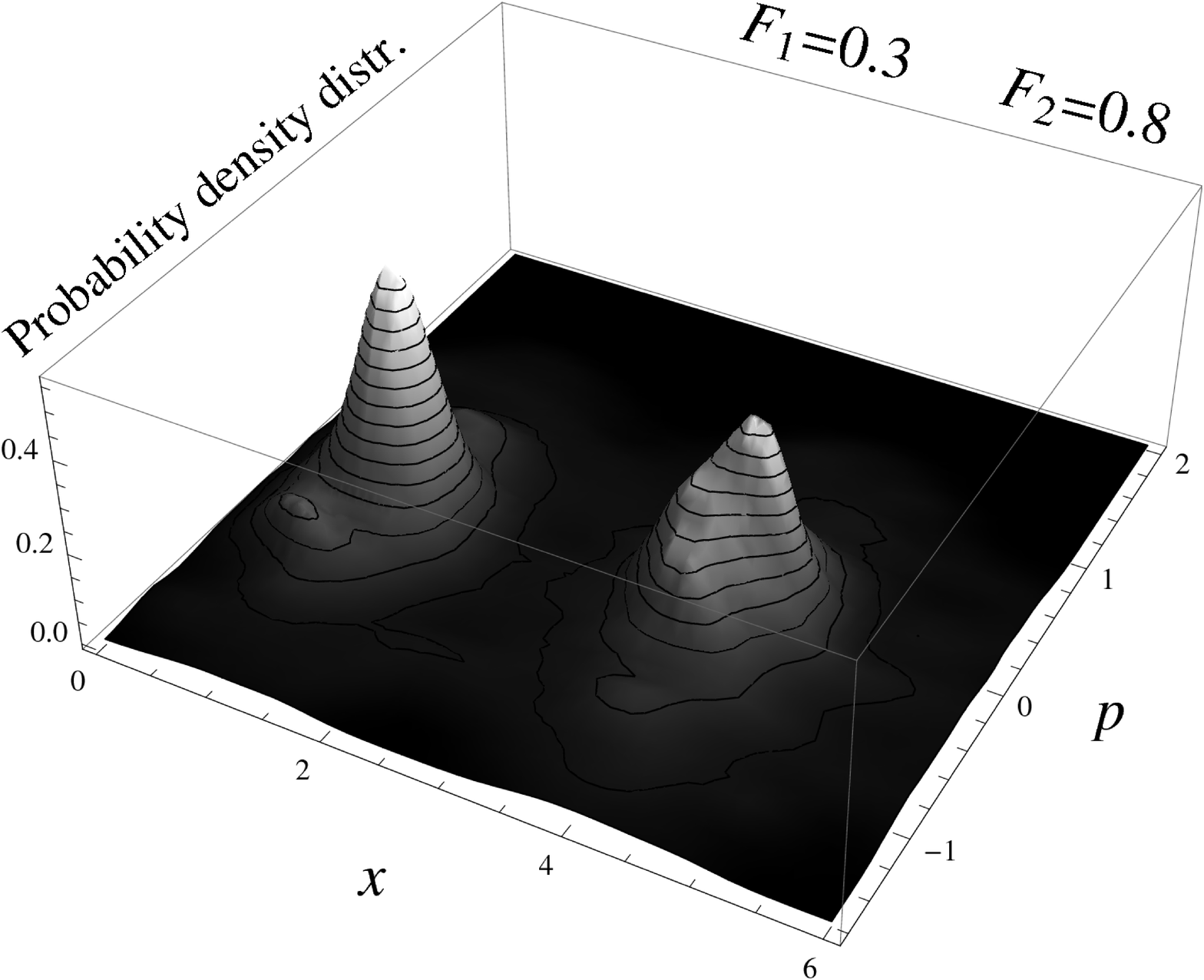}
\caption{Snapshots of the $\mu$-phase-space at $t=80$ for different parameters. The lower panels represent the probability density distribution of the corresponding 2-dimensional portrait.}
\label{water-bag-evolution}
\end{figure}

In \figref{water-bag-evolution}, the phase space is displayed at a fixed time, in this case $t=80$ (arbitrary units), for different values of the coupling constants $F_n$ (as marked on the plots). 
The phase space portraits (and the related probability density distribution) show that a fraction of particles clusters into $n$ coherent clumps which rotate according to the resonance induced by the waves while the non clustered particles are seen to be uniformly distributed between two oscillating boundaries. As it is shown below, the rotation of the clumps is synchronized with the oscillations of the $n^{th}$ harmonic intensity.

\emphr{It is important to remark that other dynamical scenarios are allowed \cite{DIIb}, where negative perturbation vortices (holes) \cite{BNR70} and also hole-clump dipoles \cite{Neg05} can spontaneously emerge from the self-consistent interaction between particles and waves. Different situations are triggered by the selected class of initial conditions and/or chosen parameters. In this respect, a crucial parameter is the pulsation $\omzn$ which appears in the Hamiltonian \reff{GWM_H} (also called the detuning parameter in the context of Free-Electron Laser studies). When setting $\omzn$ to a value different from zero (and above a critical threshold) both a hole and a massive agglomeration appear in the phase space \cite{A05}. At variance, in our numerical experiment, $\omzn$ is set to zero, disappearing in \ereff{eq555} and one (or, alternatively, two) filamented macro-particle(s) is (are) found. This is consistent with the results reported in \cite{A05} and in \cite{TMM94} where the dynamics in the phase space (for the limiting case where one wave is solely allowed for) is explained in terms of a reduced mathematical model (see also \cite{AF06}). In principle, it should be possible to extend this approach by invoking an ad hoc generalization of the macro-particle concept to tackle the regime where the clump-hole dipole sets in. Specifically, one could ideally describe the system in terms of two macro-particles, one associated to the hole and the other to the clump.}

Numerical simulations of the $E_n$ evolution described by the system \reff{EqMotionSim}  show how the amplification of the waves is characterized by several subsequent steps. First of all, the values of the coupling constants select the leading harmonic.
\begin{figure}[!ht]
\centering
\includegraphics[width=0.45\hsize]{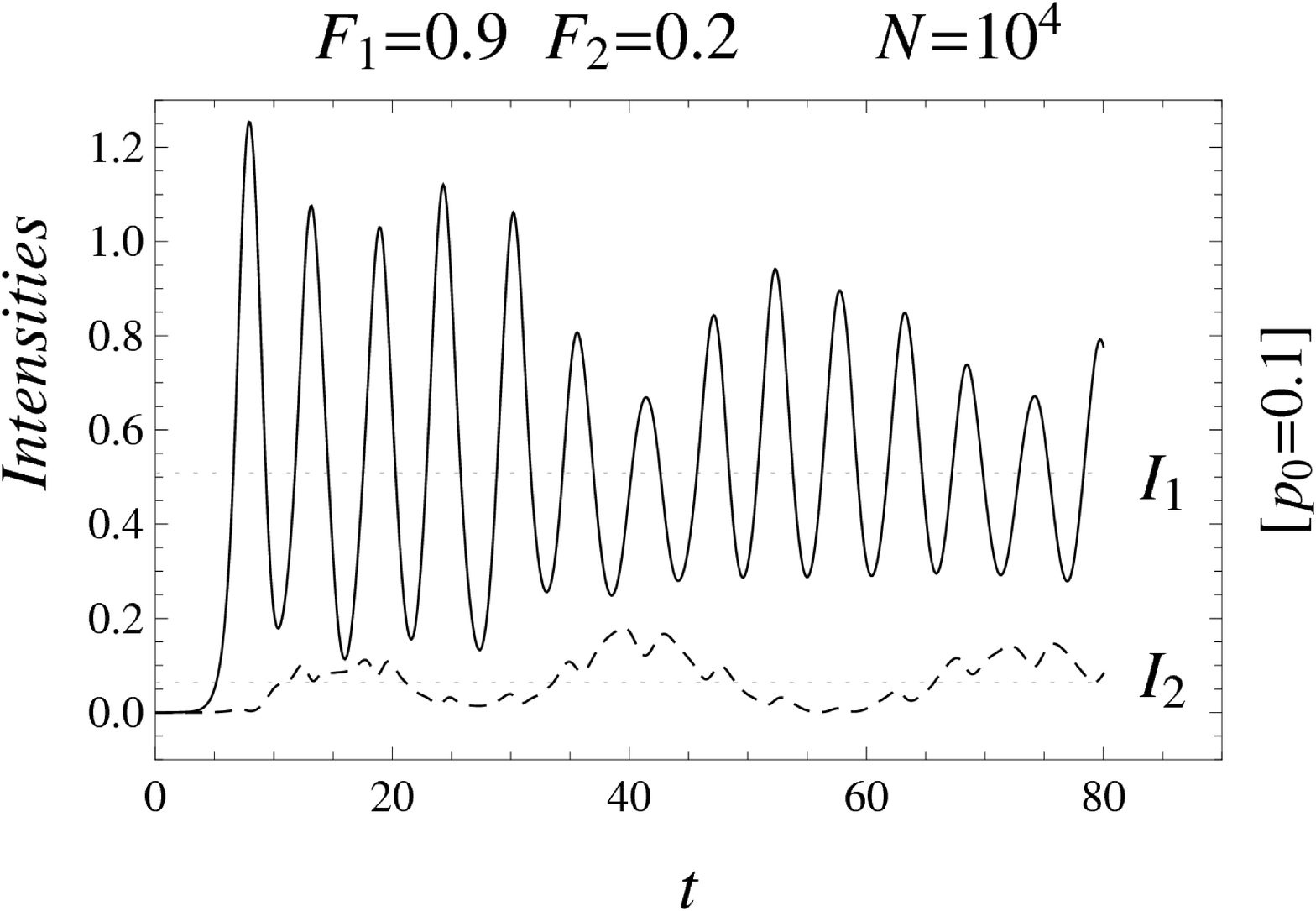}
\includegraphics[width=0.45\hsize]{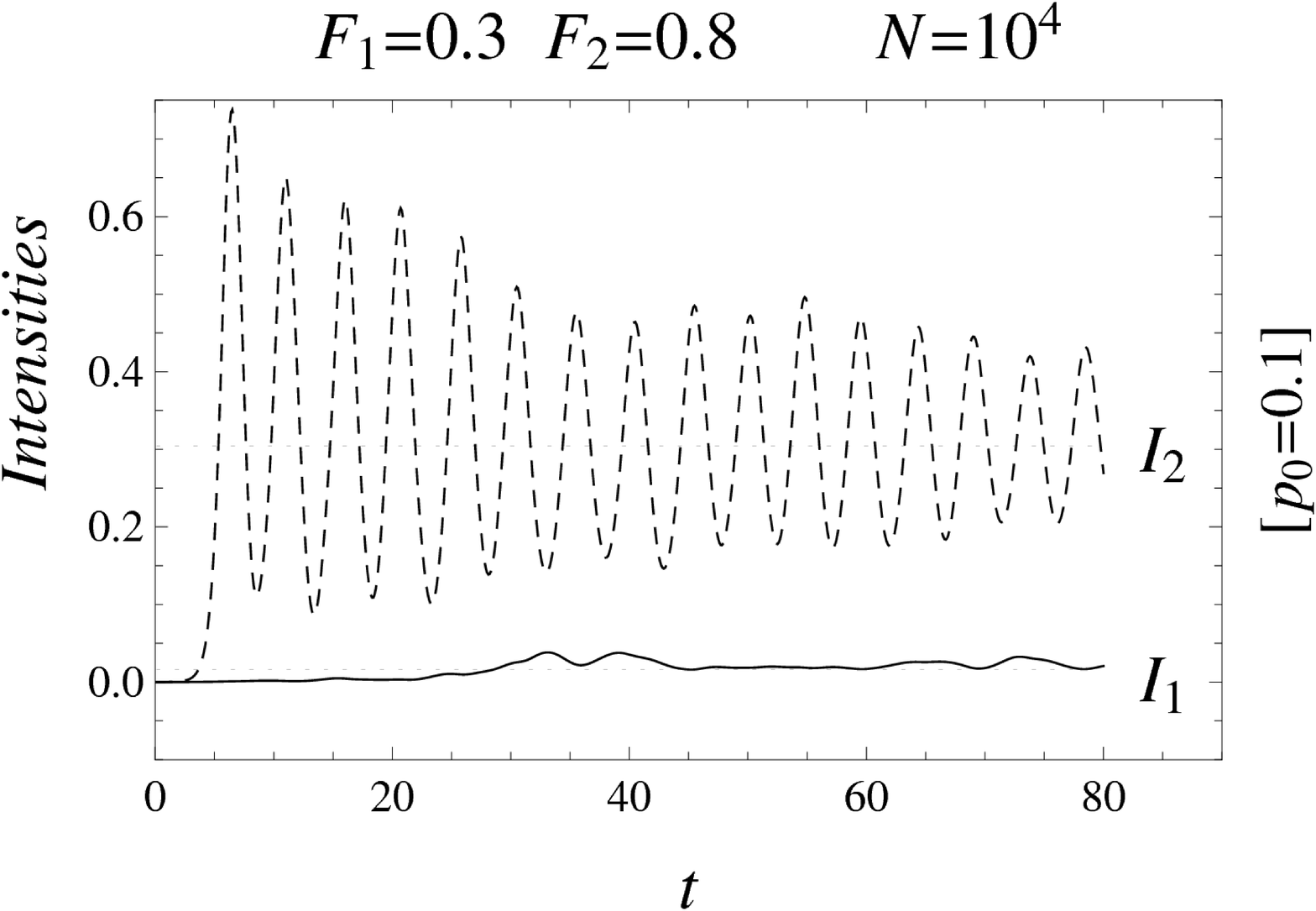}
\caption{Plots of the Langmuir-wave intensities. The dotted lines denote the mean values of $I_n$.}
\label{intensities}
\end{figure}
Initially the amplitude of this mode exponentially grows in time depending on the initial conditions of the electron beam injected into the plasma. In a typical situation (see \figref{intensities}), after this exponential instability, the intensity of the $n^{th}$ leading harmonics begin to oscillate around some average value. We observe that the present analysis differs from the linear Landau damping effects \cite{LP81}, where the plasma-wave interaction and the resulting instabilities are regulated by the slope of the unperturbed initial distribution function at the resonance condition. Despite here the distribution is homogeneous in the phase space, unstable modes arises as result of the non-linear interaction of the beam-plasma system \cite{B04}. In this sense, we are analyzing the so-called non-linear Landau damping, in which full non-perturbative effects are taken into account (for a basic study of the instabilities in the case of uniform initial distribution function, see \cite{B84}).

\subsection{Thermalization and resonance}\label{RES}
In what follows, we focus on a phenomenological analysis of the resonance band of the model by studying how the increase of the \emphr{beam-particle initial momentum spread} affects the dynamics of the system once the model parameters $F_{1,2}$ are fixed. In particular, it is found that a transition phenomenon occurs between qualitatively different dynamic states. In other words, \emphr{it is found a threshold value of $p_0$ (defined in \ereff{initialcond})}, such that below this threshold the system responds to the beam injection by clustering the beam particles into the resonant rotating clumps.
\begin{figure}[!ht]
\centering
\includegraphics[width=0.60\hsize]{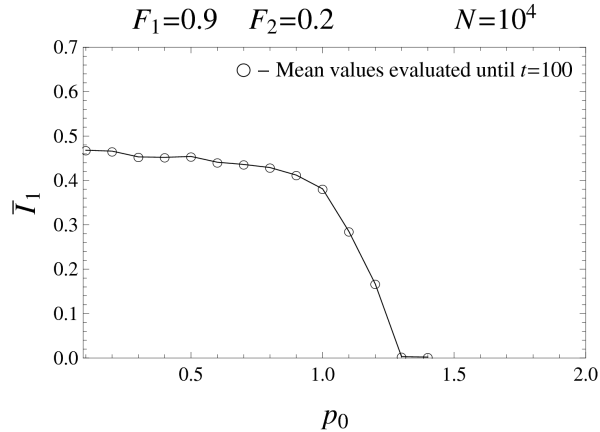}
\caption{Plots of the mean intensity $\bar{I}_1$ in function of $p_0$ defined in \ereff{initialcond}. Each symbol represents the mean value averaged for $0<t<100$ calculated by a single simulation with $N=10^{4}$.}
\label{I1vsP0}
\end{figure}
Numerical simulations are performed by  considering, as initial conditions (we recall that the beginning of the evolution is taken at $t=0$), $N=10^{4}$ particles randomly distributed in the phase-space defined by \ereff{initialcond}, and by varying $p_0$ in the region $0<p_0<1.6$. Each simulation is performed until $t=100$ and the mean value of the intensity $\bar{I}_1$ (we set $F_1=0.9$ and $F_2=0.2$) is plotted in \figref{I1vsP0} as function of $p_0$.

\begin{figure}[!ht]
\centering
\includegraphics[width=0.35\hsize]{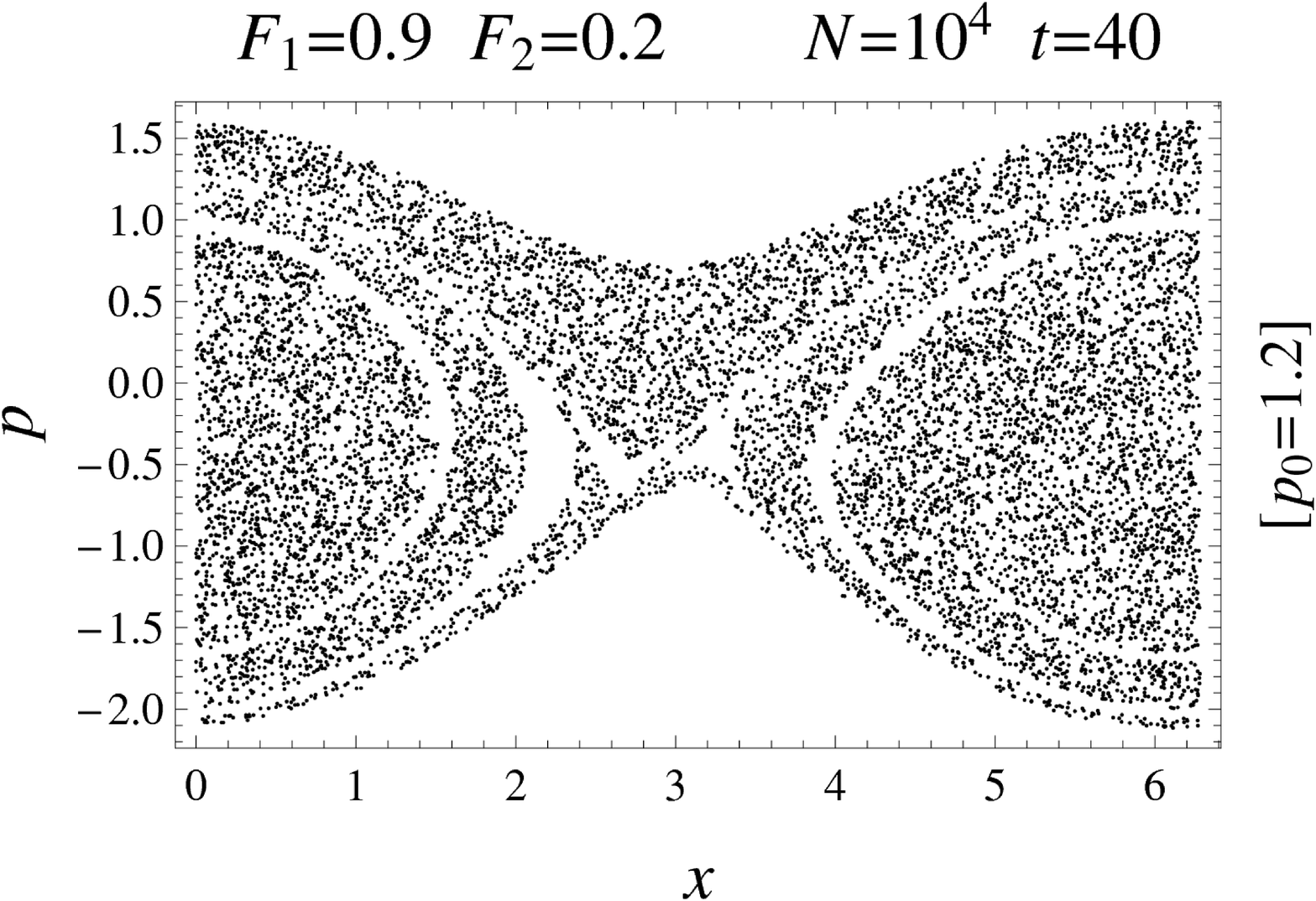}
\includegraphics[width=0.35\hsize]{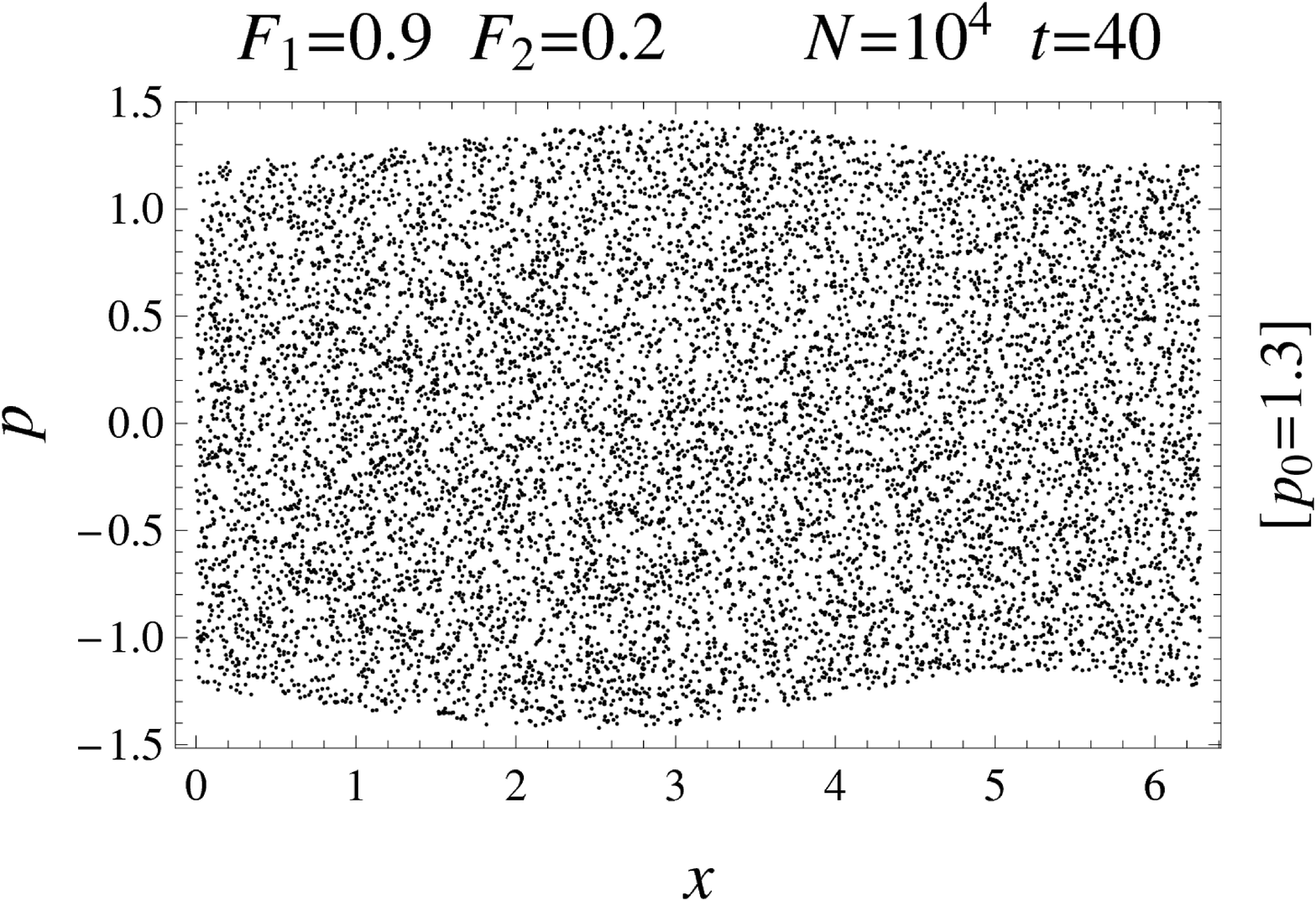}
\caption{Snapshots of the $\mu$-phase-space  at $t=40$ for two different initial conditions: $p_0=1.2$ (left panel) and $p_0=1.3$ (right panel).}
\label{NoRes}
\end{figure}

As a result a sharp transition is found between the resonant and non-resonant regimes. In particular, for $p_{0}\gs1$, the thermalization of the beam starts to be much less effective until the value $p_0\simeq1.3$ is attained, where the system does not respond at all and the particles are not trapped by the wave. In this case, no increase of the intensity is observed, since the particles do not resonantly rotate, as it can be argued from \figref{NoRes}, where the response of the system for $p_0=1.2$, $1.3$ is reported. This analysis has been carried out for values of the parameters $F_{1,2}$ corresponding to the response of the second harmonic and the results are qualitatively comparable. \emphr{As a remark, we underline that it is straightforward to verify that the threshold on $p_0$ is present also for the case of one single wave. In particular, running the simulations by setting $F_1=1$ and $F_2=0$, the dynamical effect of one leading harmonic results in a positive shift of about $0.2$ of the threshold value for $p_0$ (for the same conditions of \figref{I1vsP0}).}

\begin{figure}[!ht]
\centering
\includegraphics[width=0.35\hsize]{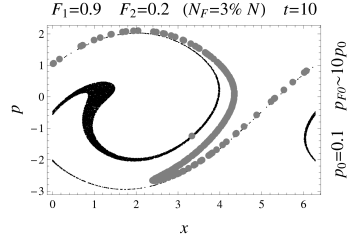}
\caption{At time $t=10$ and  $F_1=0.9$, $F_2=0.2$ and $y=10$ ultra-fast particles (marked in gray with large-size points) are involved in the dynamics.}
\label{thermal-time}
\end{figure}

In order to better characterize the resonance behavior of the system, we now consider how the threshold previously obtained can affect the thermalization properties of a \emphr{small population of beam particles identified by large initial momenta.} Numerical simulations have been performed by considering, as initial conditions, $N=10^{4}$ \emphr{beam particles} divided in two distinct populations. Accordingly, the \emphr{most part of the beam} (labeled with the subscript ``$B$'') is randomly distributed in the region $-p_{0}<p_{B}<p_{0}$, with $p_0=0.1$, while the ultra-fast population (labeled by ``$F$'' and marked in gray with large-size points in all plots reported here) is set as $N_F=0.03 N$ and is randomly distributed in the region $-\pi<x<\pi$ and $y p_{0}<p_{F0}<yp_{0}+10^{-2}\,yp_{0}$ (for simplicity, we indicate $p_{F0}\sim yp_{0}$ over the graphs). \emphr{We remark that such a population has been treated as belonging the beam, \ie we describe a single beam composed by a small percentage of extra-fast constitutive elements}.
\begin{figure}[!ht]
\includegraphics[width=0.35\hsize]{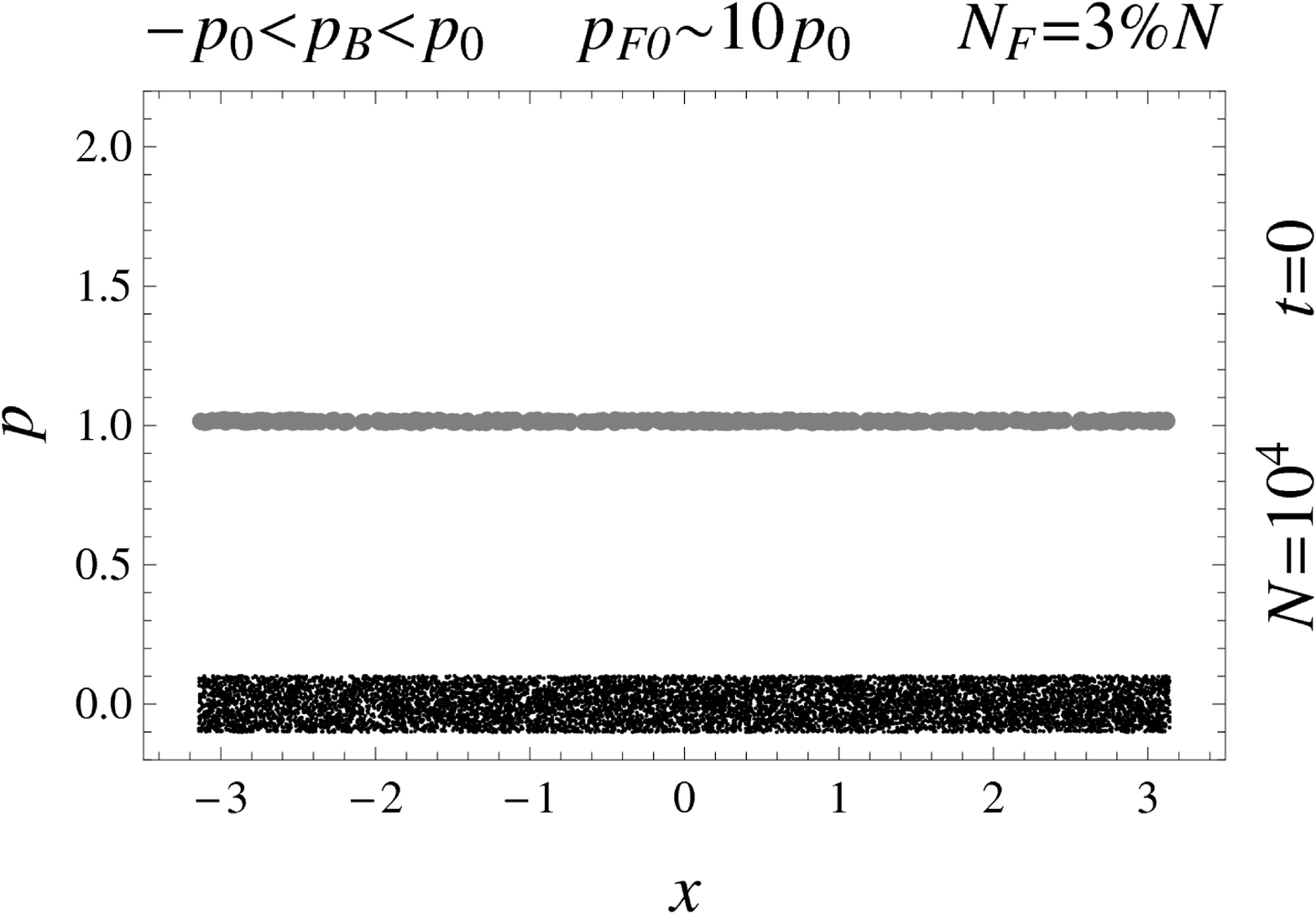}
\includegraphics[width=0.35\hsize]{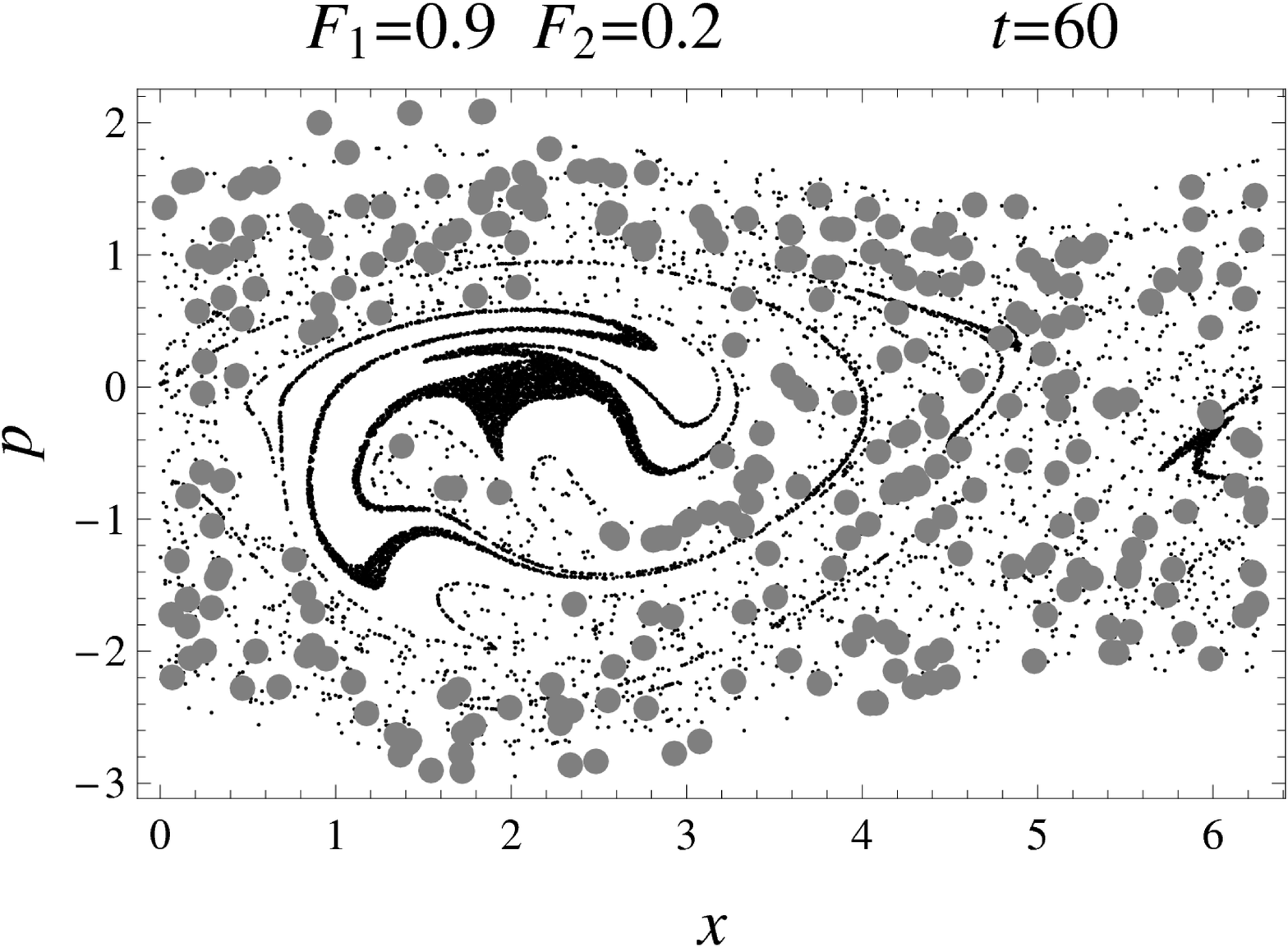}
\includegraphics[width=0.27\hsize]{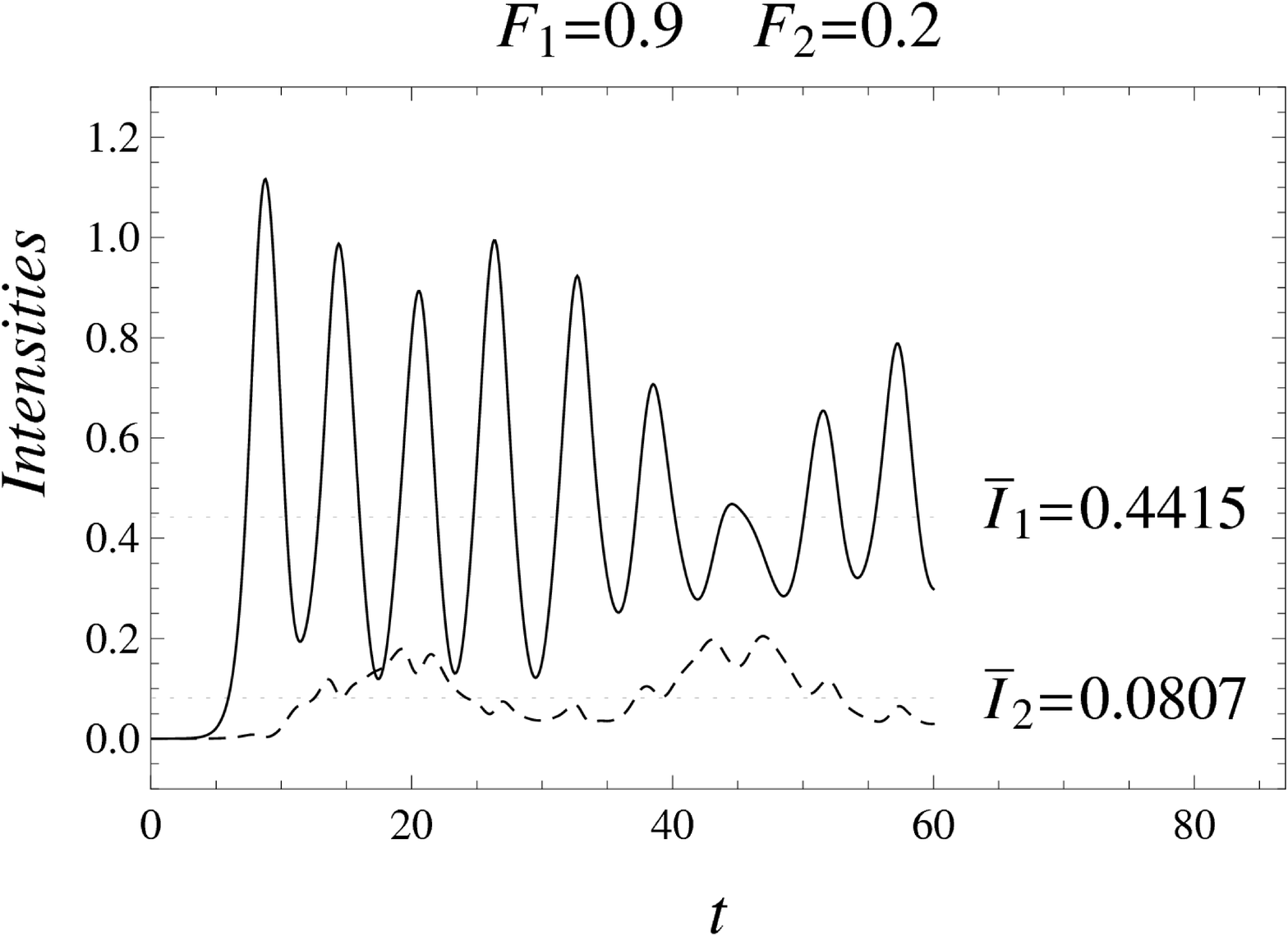}
\vspace{1mm}\\
\includegraphics[width=0.35\hsize]{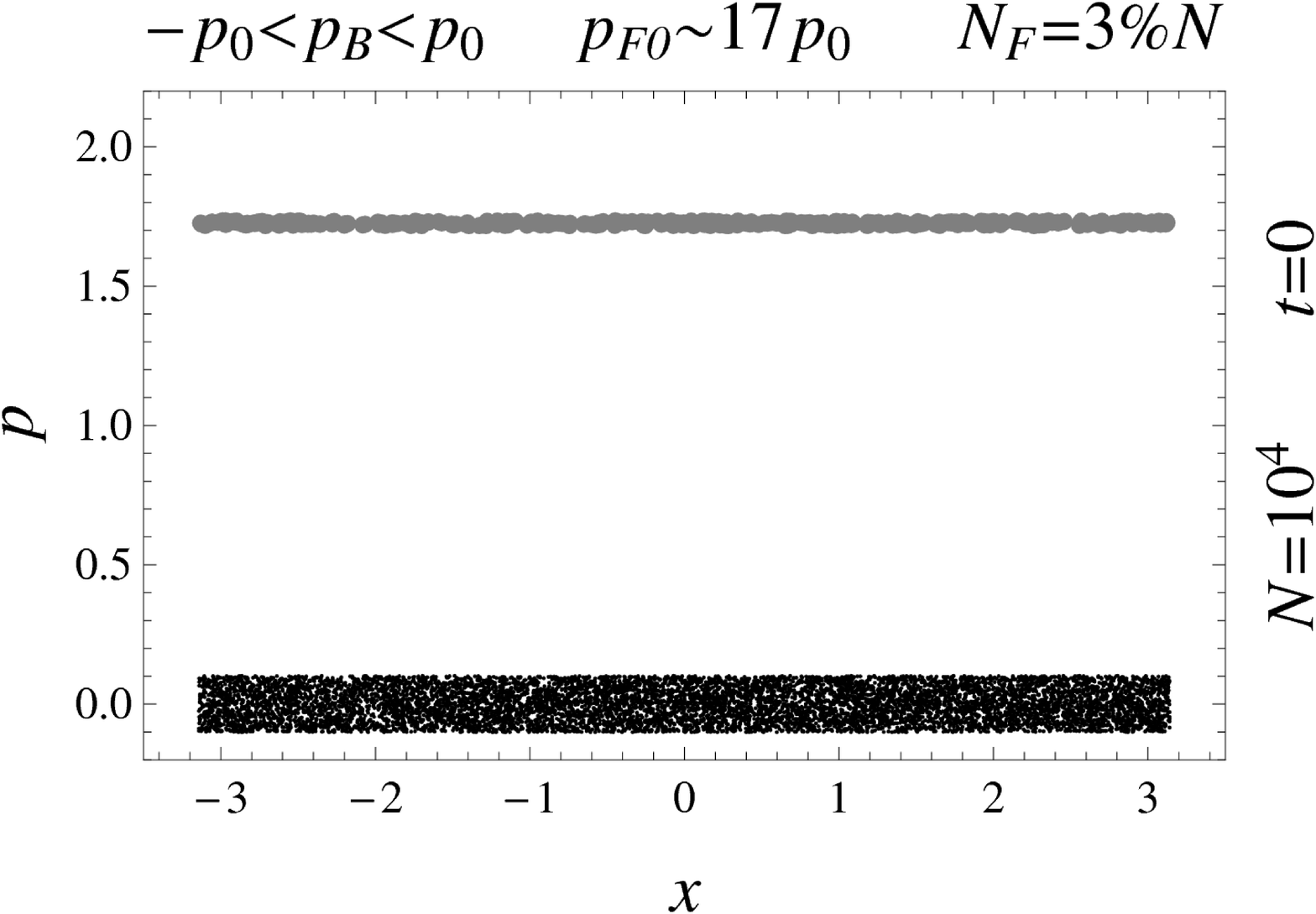}
\includegraphics[width=0.35\hsize]{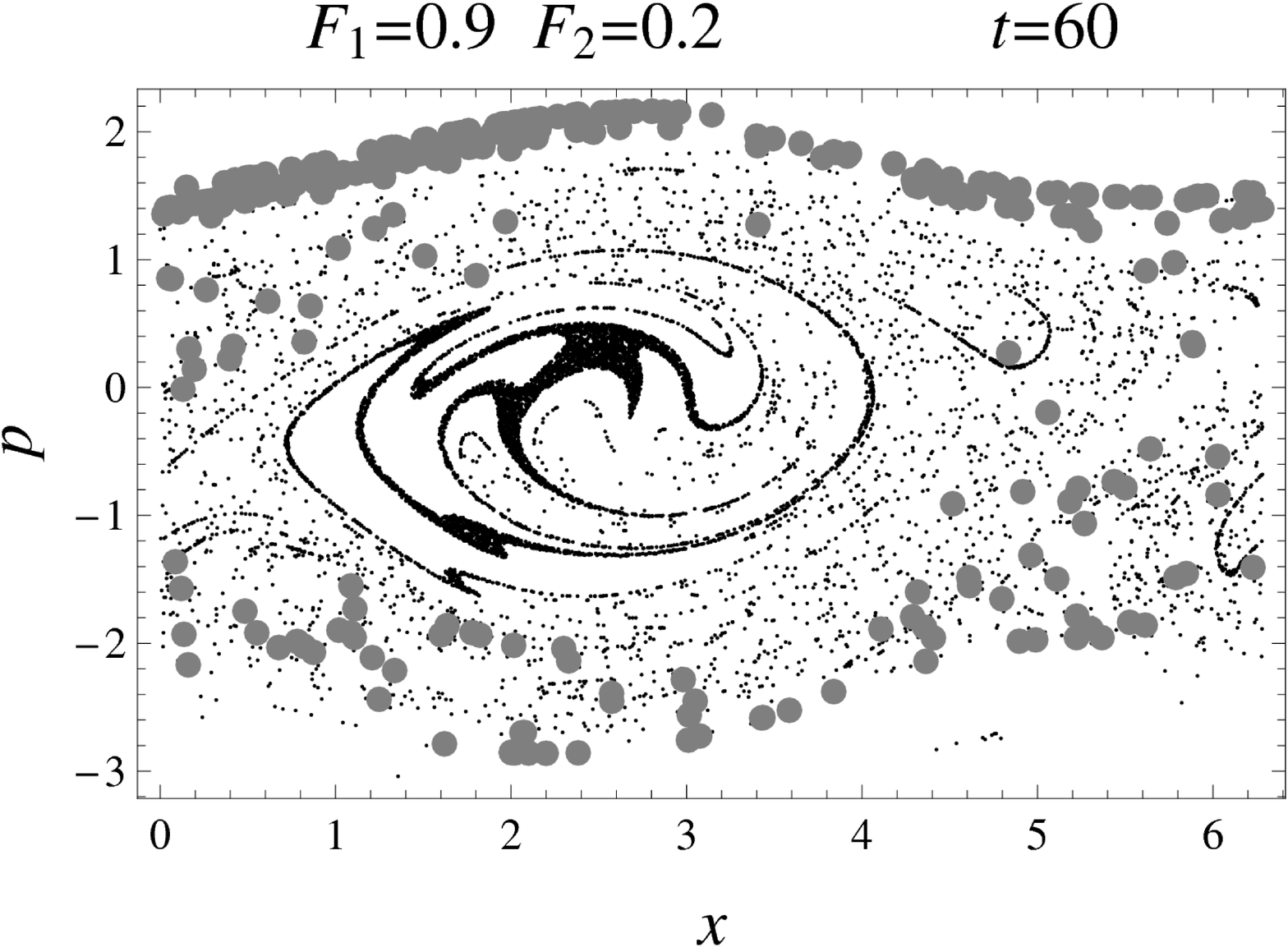}
\includegraphics[width=0.27\hsize]{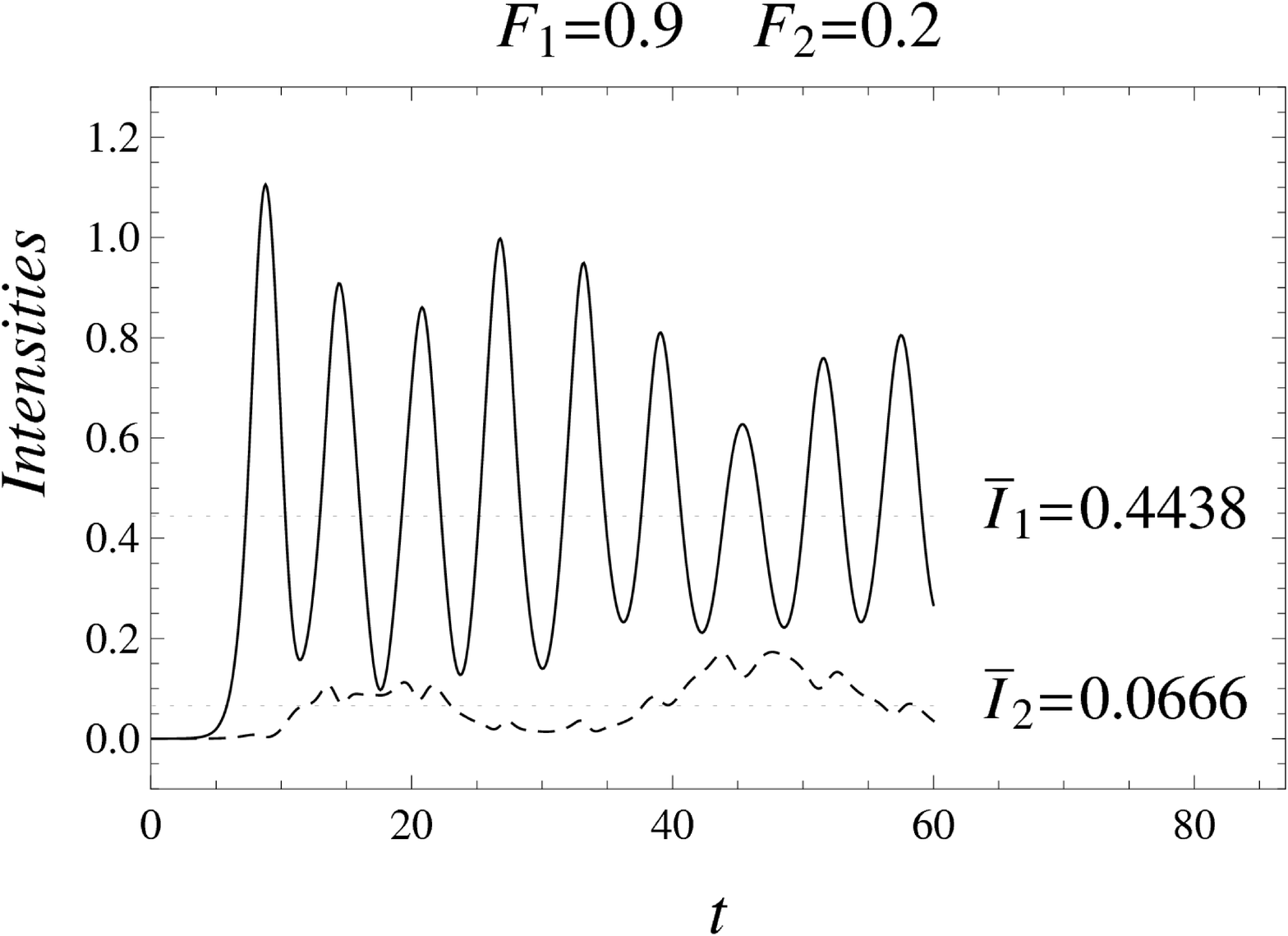}
\vspace{1mm}\\
\includegraphics[width=0.35\hsize]{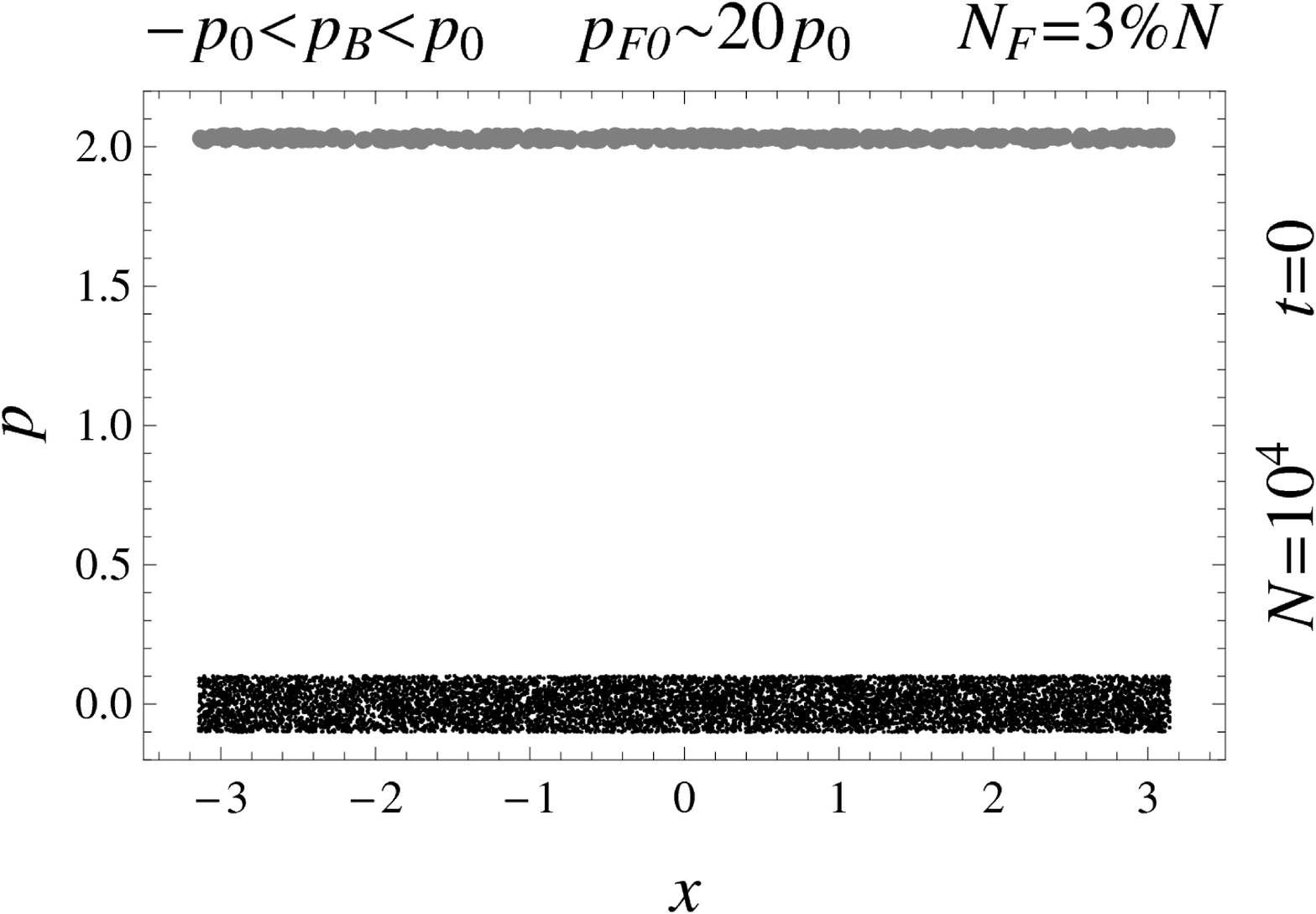}
\includegraphics[width=0.35\hsize]{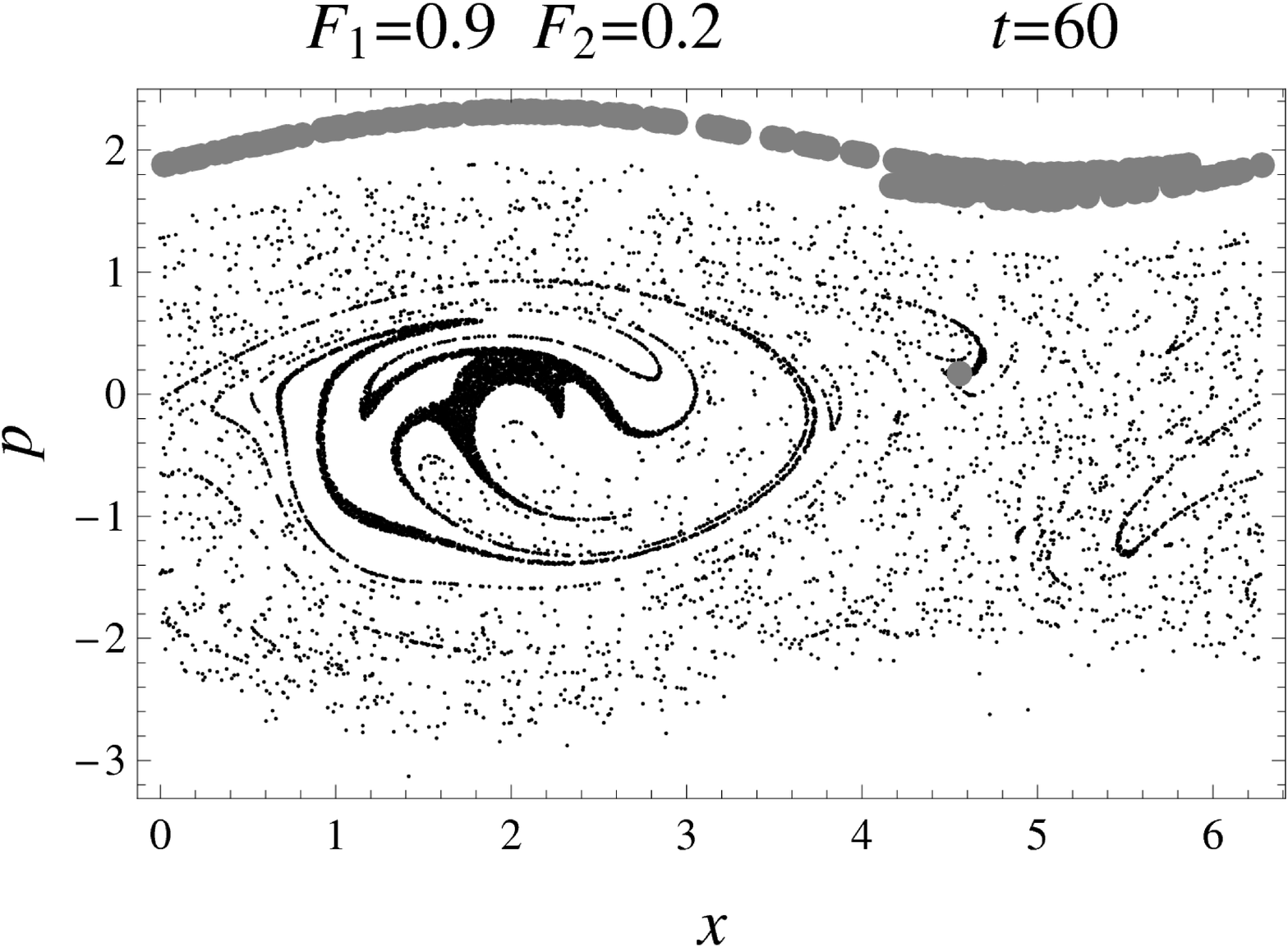}
\includegraphics[width=0.27\hsize]{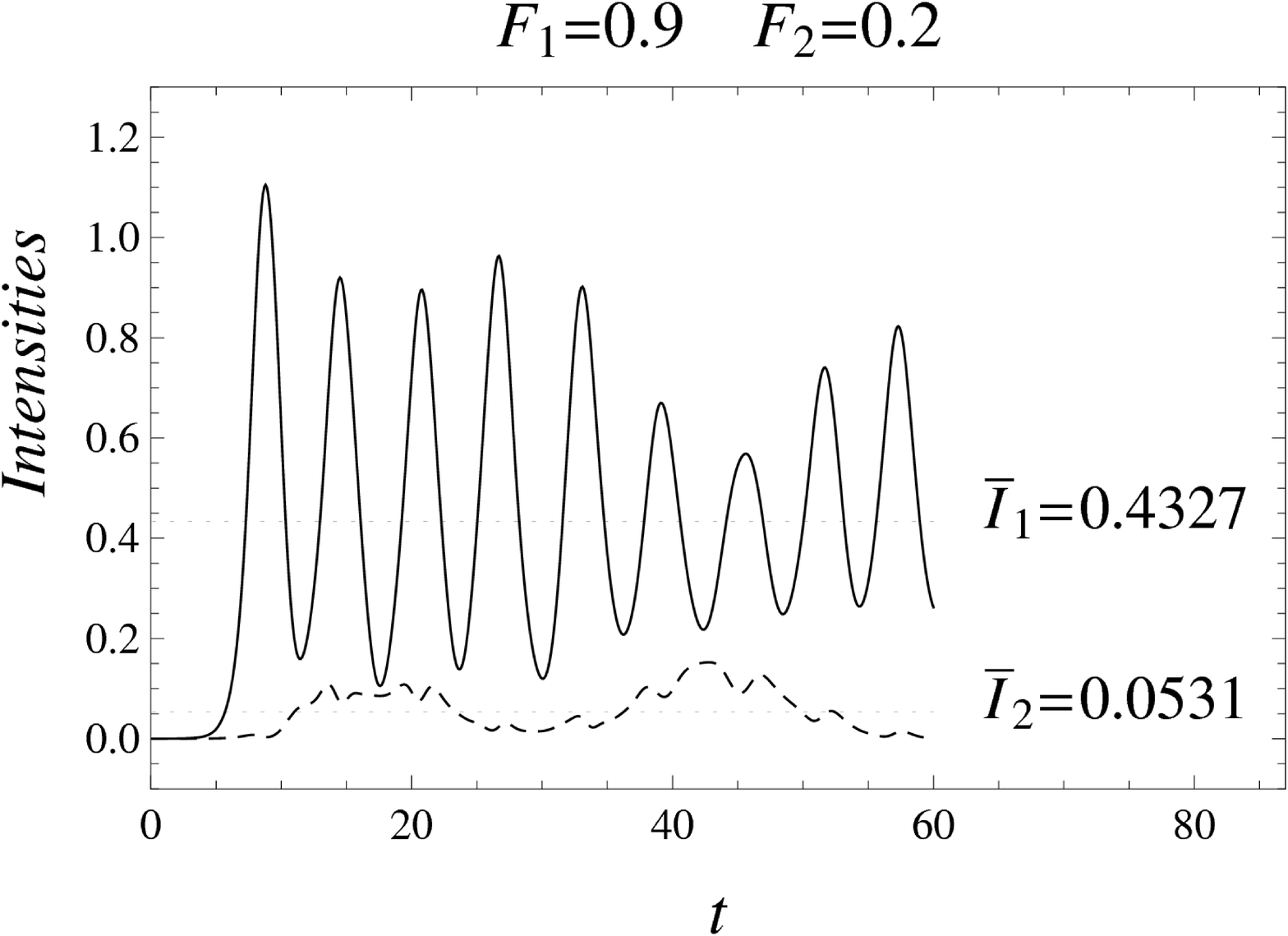}
\caption{Simulations for $F_1=0.9$, $F_2=0.2$ and $y=10,\;17,\;20$. The extra population is marked in gray with large-size points.}
\label{1-x}
\end{figure}
The outcomes of the simulations are displayed for different values of the factor  $y$ and phase space snapshots are given at a fixed time (in all cases, $t=60$) for different coupling constants $F_n$ (as indicated above  the graphics frames). In \figref{1-x}, some examples are reported of simulations performed by setting the values of the coupling constant in the region of the parameter space fixed by $F_1=0.9$ and $F_2=0.2$. The plots correspond to different values of $p_{F0}\sim yp_{0}$. In particular, we have set $y=10,\;17,\;20$, respectively. In \figref{1-x}, the physical effect of the expected threshold in $p_{F0}$ -- that discriminates whether the thermalization of the extra population occurs -- is evident. In particular, for $y\ls20$ the particles are not far enough from resonance and, during the evolution of the system, they mix with the whole beam in the collective dynamics. On the other hand, for $y\gs20$ the fast particles remain isolated from the clusters and do not resonate with the waves. At variance, \figref{thermal-time} shows that for $y\ls20$ (namely, $y=10$) the ultra-fast particles behave like the beam particles - thus ``thermalizing'' with these ones - on a time scale of the order of $t\ls10$. This indicates how the ultra-fast beam participates in the initial exponential amplification of the wave amplitude.

\begin{figure}[!ht]
\includegraphics[width=0.35\hsize]{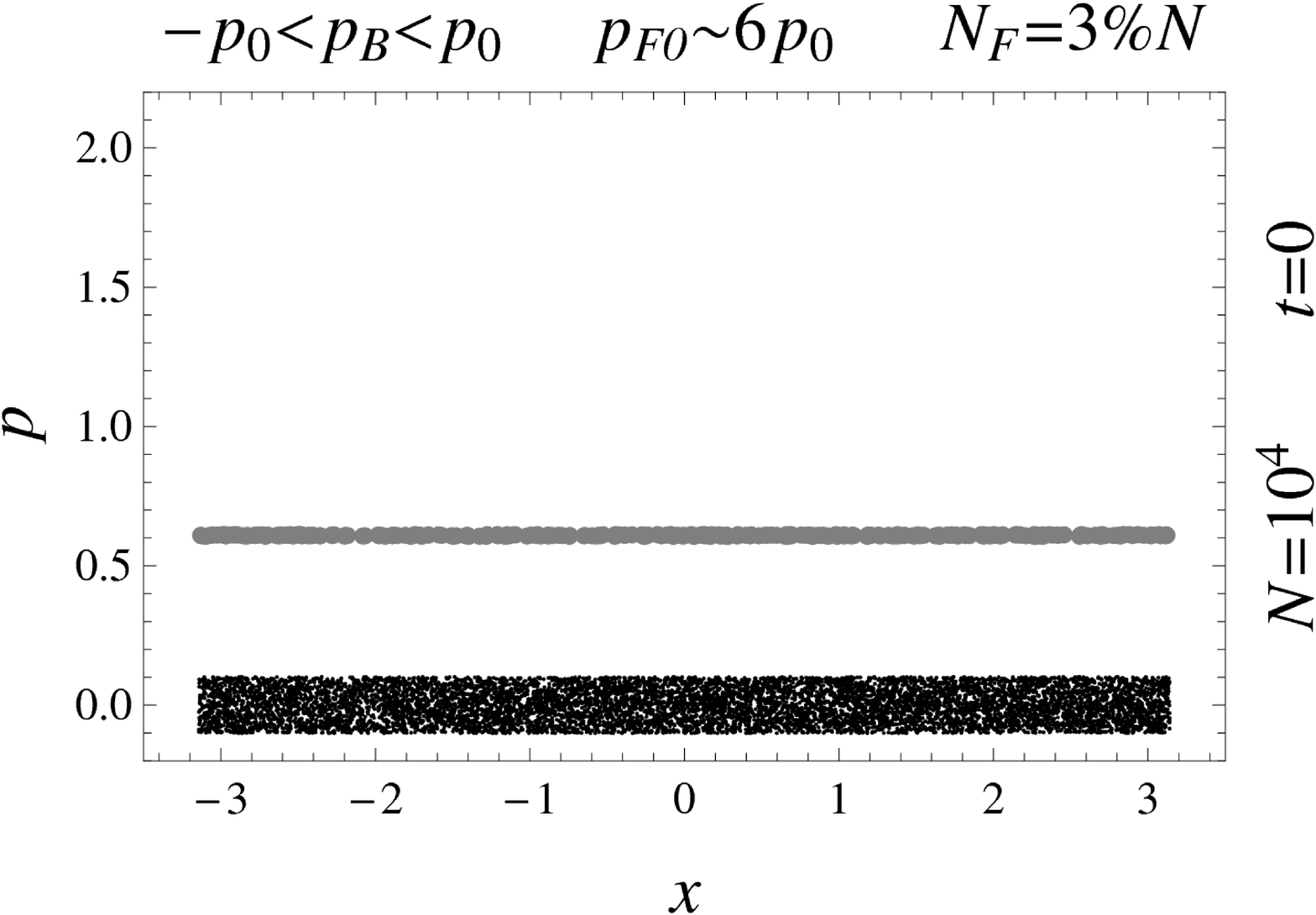}
\includegraphics[width=0.35\hsize]{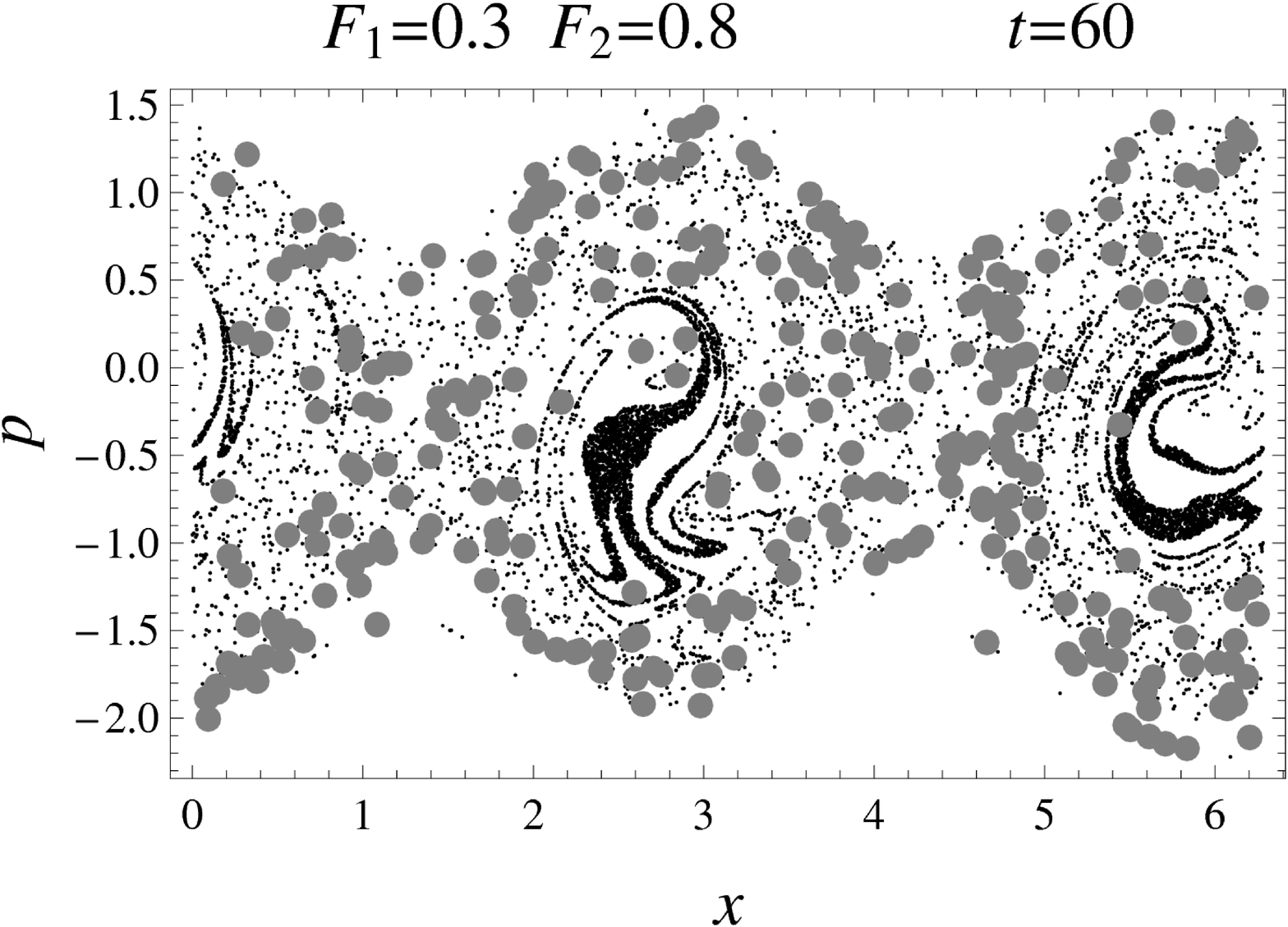}
\includegraphics[width=0.27\hsize]{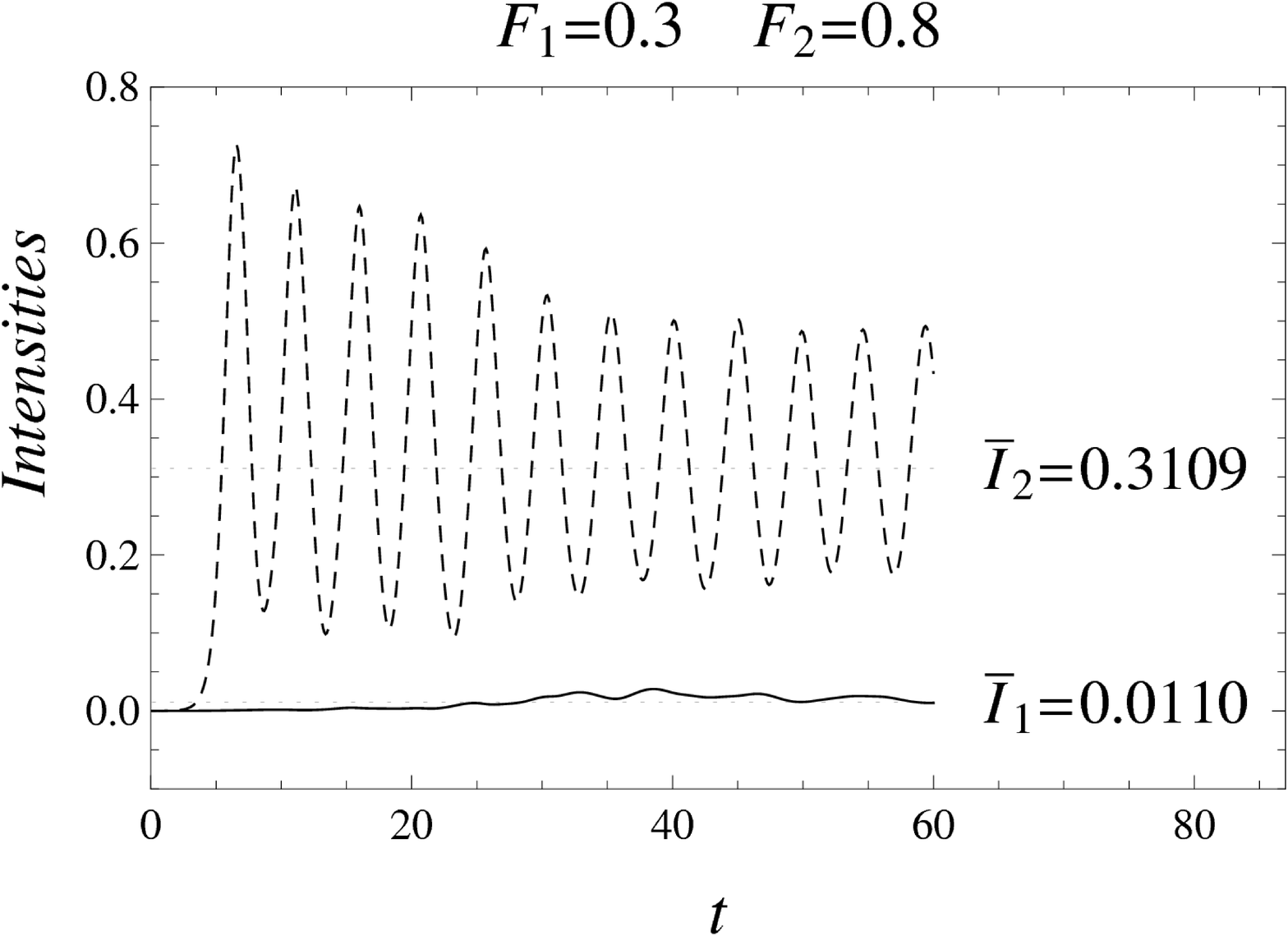}
\vspace{1mm}\\
\includegraphics[width=0.35\hsize]{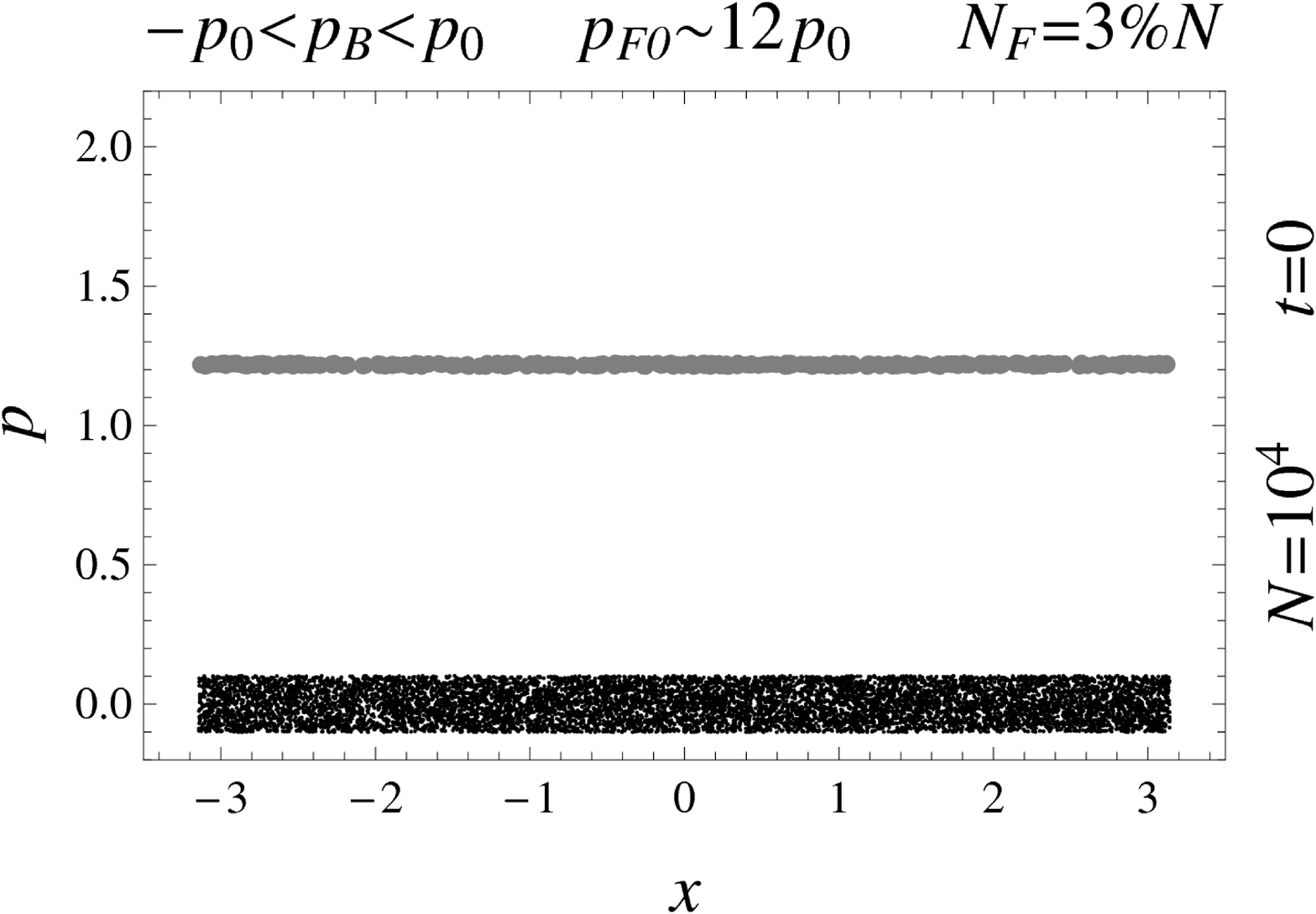}
\includegraphics[width=0.35\hsize]{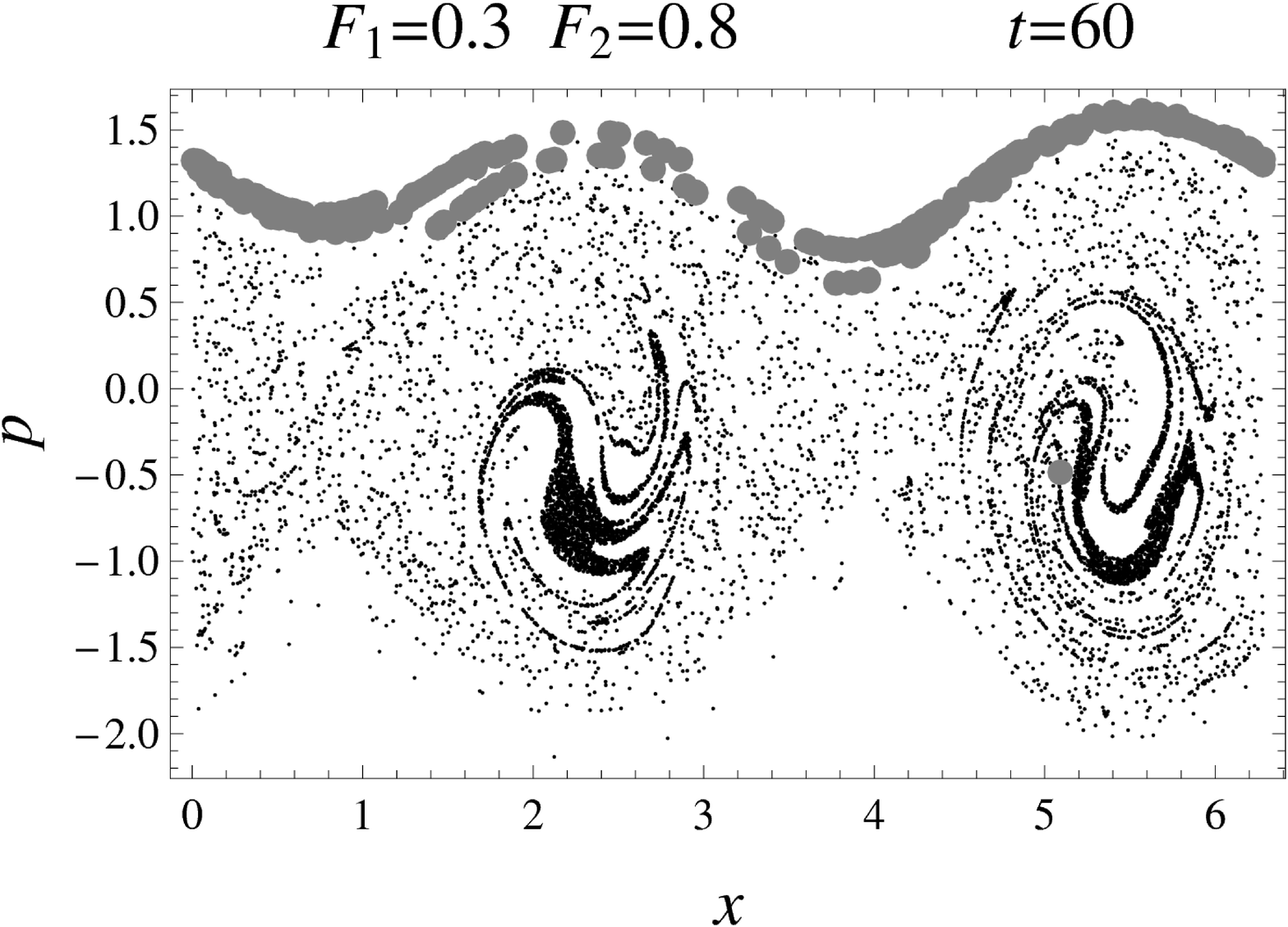}
\includegraphics[width=0.27\hsize]{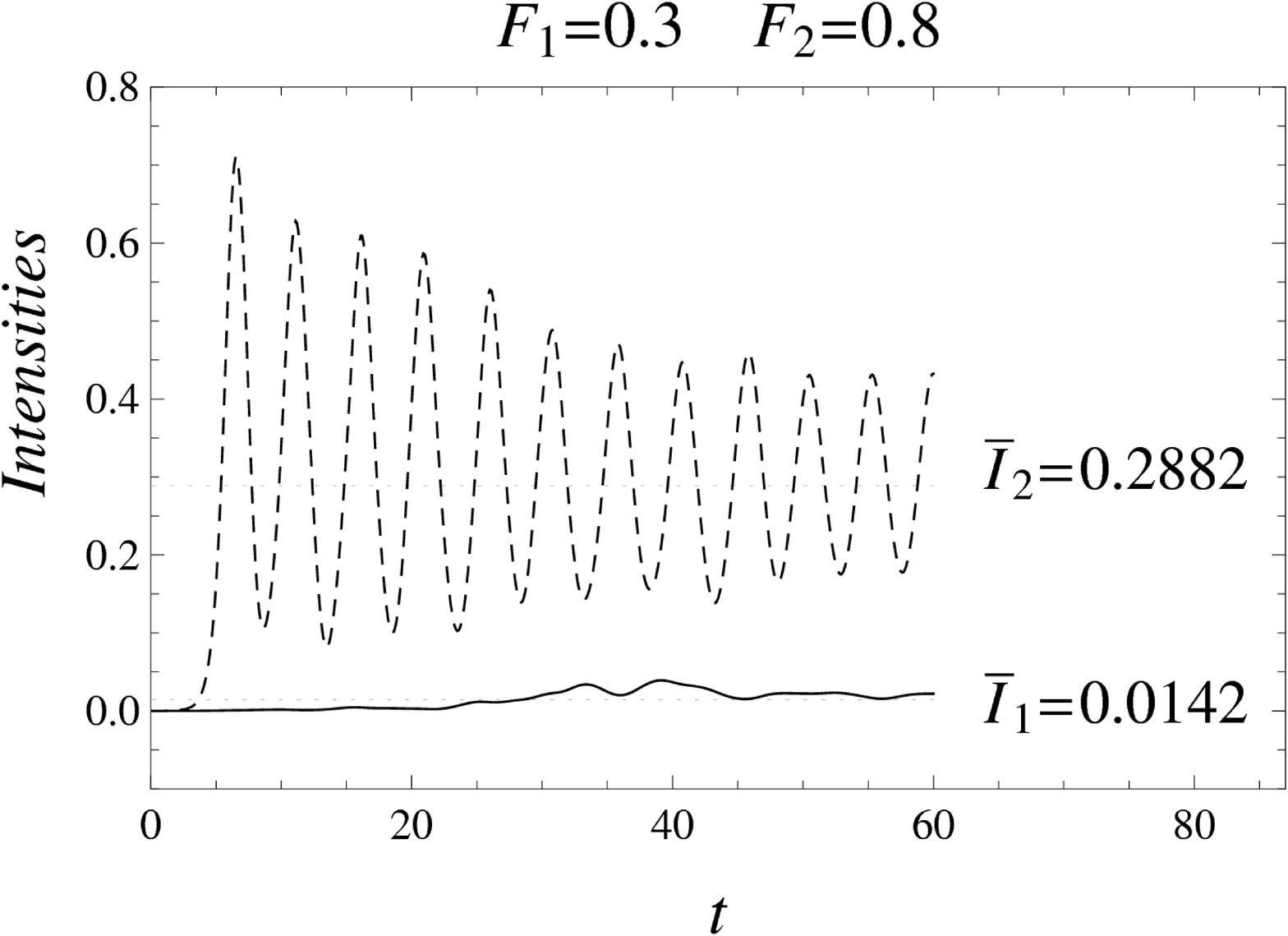}
\caption{Simulations for $F_1=0.3$, $F_2=0.8$ and $y=6,\;12$. The extra population is marked in gray with large-size points.}
\label{2-x}
\end{figure}

In view of this analysis, the resonance threshold described in \figref{I1vsP0} appears as possibly underestimated. In particular, the extra-fast particle population can be regarded as a momentum marker that probes the thermalization of the system. This result can be extended also to the case in which the leading harmonics is the second one, a condition obtained by fixing the coupling constants as $F_1=0.3$ and $F_2=0.8$. The plots in \figref{2-x} correspond to $y=6,\;12$, respectively. The threshold value for the $p_{F0}$ is evident also for this parameter choice and we find that for $y\ls12$ the extra fast population thermalizes.

Let us now consider the effect produced by an initial localization of these particles near the resonant region of the phase space. In particular, by localizing a very small population ($N^{*}=0.05 N$) with $p^{*}\simeq0.1$, a very small increase of the fully resonant particles number  is enough to obtain a mixing response of the system also in the case of out-of-resonance initial conditions.
\begin{figure}[!ht]
\centering
\includegraphics[width=0.45\hsize]{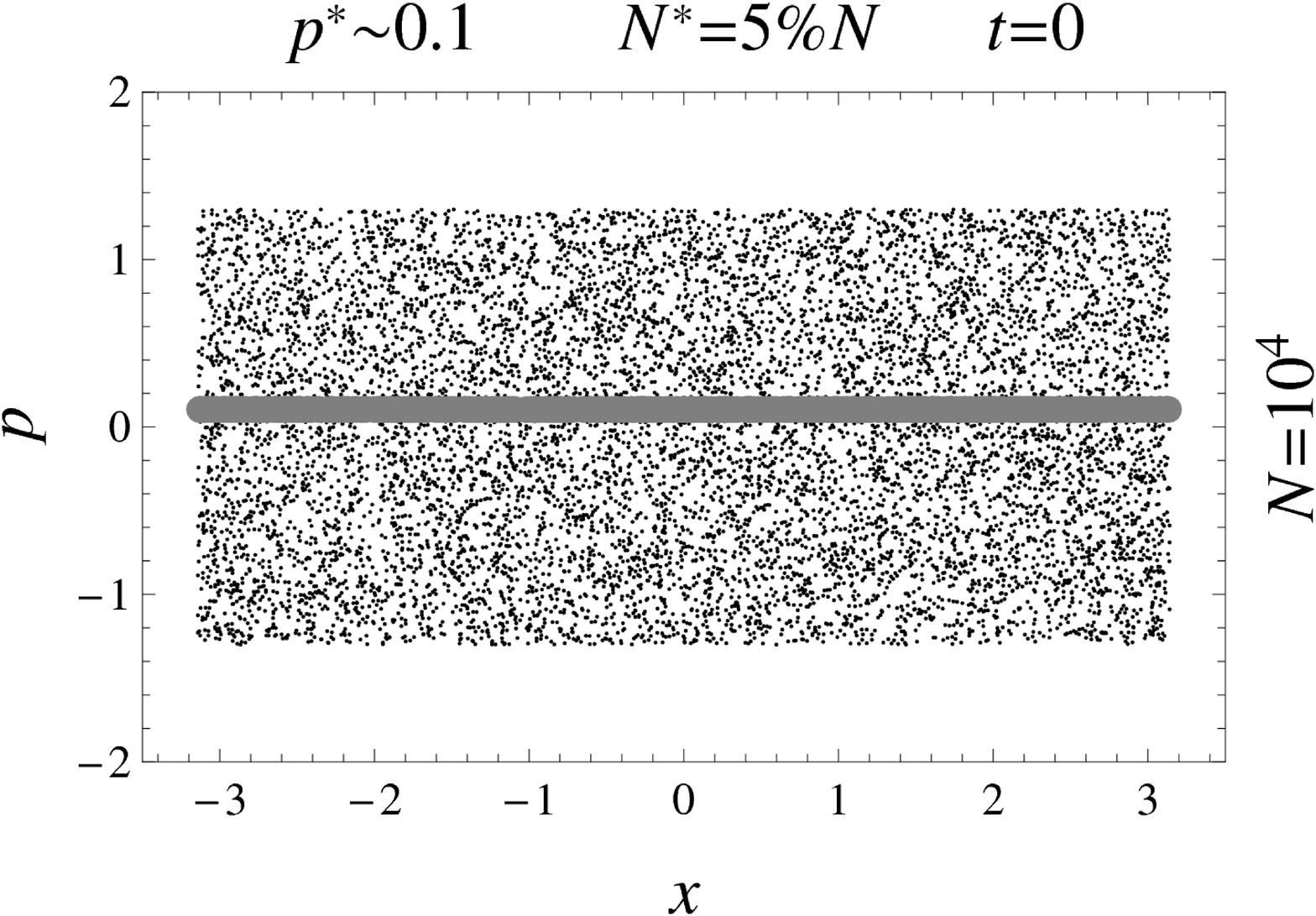}
\includegraphics[width=0.42\hsize]{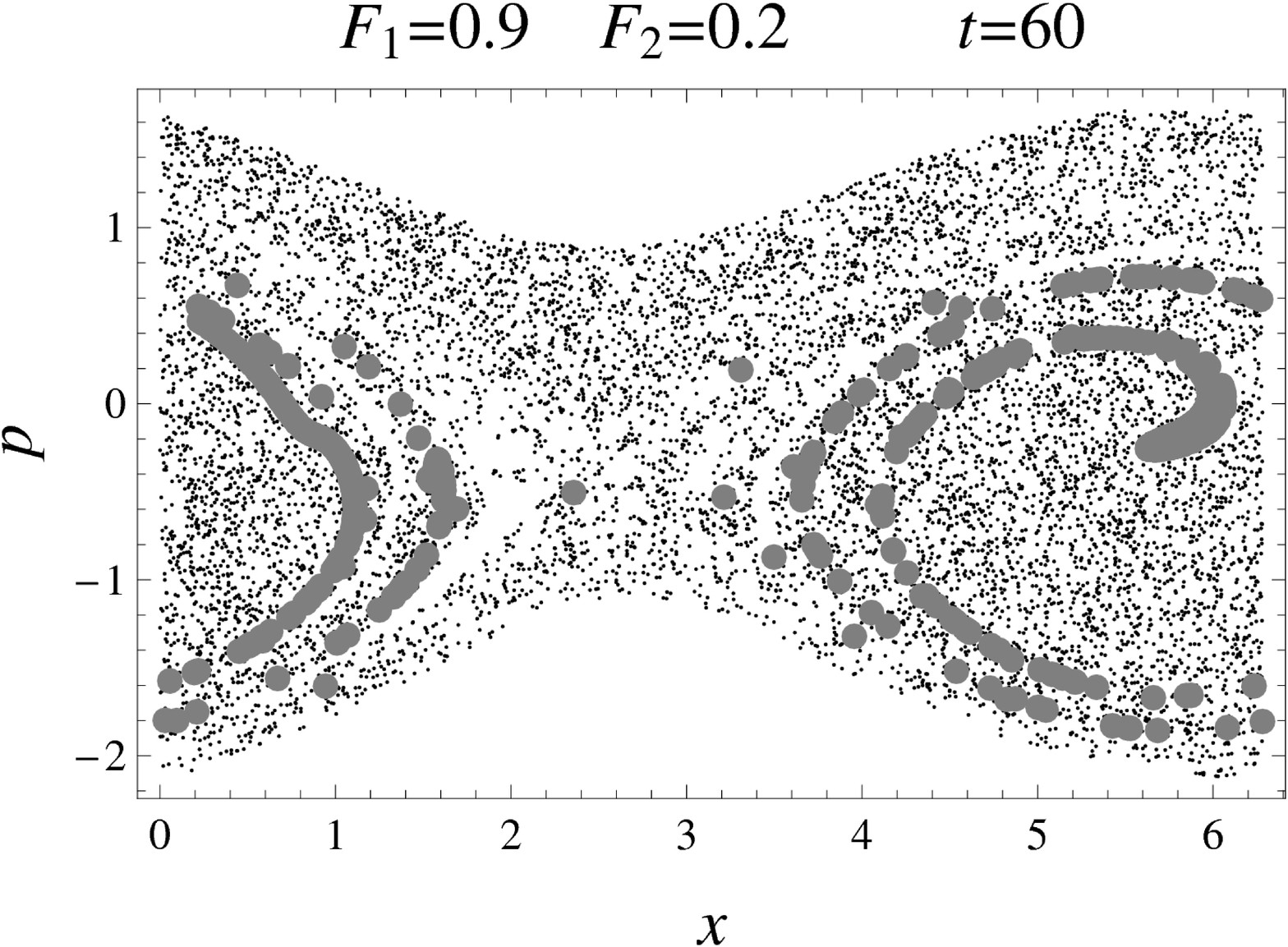}
\caption{Snapshots of the evolution of the $\mu$-phase-space particle distribution in the presence of a small ($N^{*}=0.05 N$) localized population (marked in gray with large-size points) at $p^{*}\simeq0.1$.}
\label{NoResStar}
\end{figure}
This result clearly emerges when comparing the right panels of \figref{NoRes} and \figref{NoResStar}, which are both characterized by $p_0=1.3$. This behavior is the most intriguing feature of our analysis. In fact, it is found that a small fraction of  resonant particles within a broad non-resonant beam is sufficient to activate the resonant process.

\subsection{Quasi-stationary states}
\begin{figure}[!ht]
\centering
\includegraphics[width=0.45\hsize]{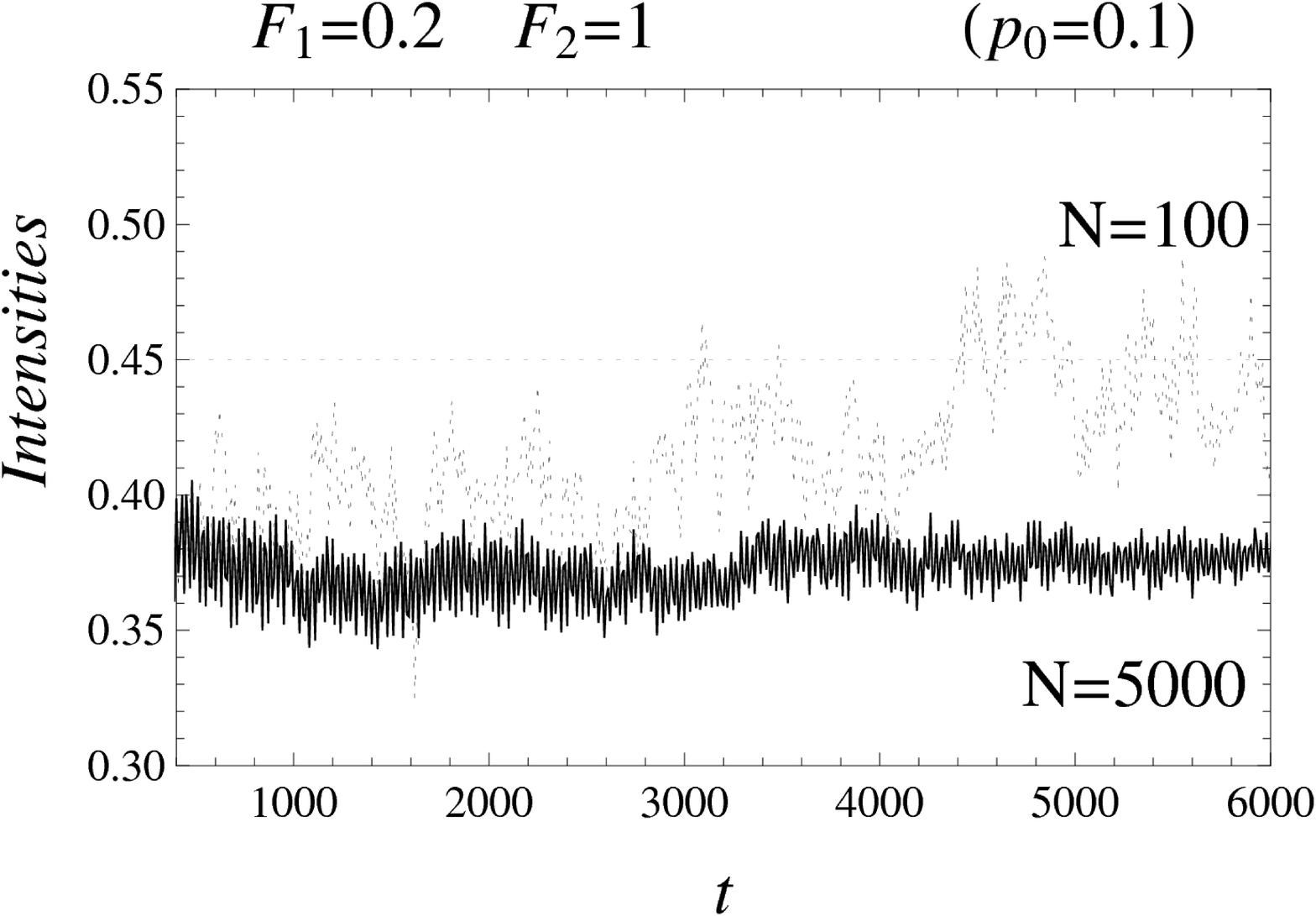}
\includegraphics[width=0.45\hsize]{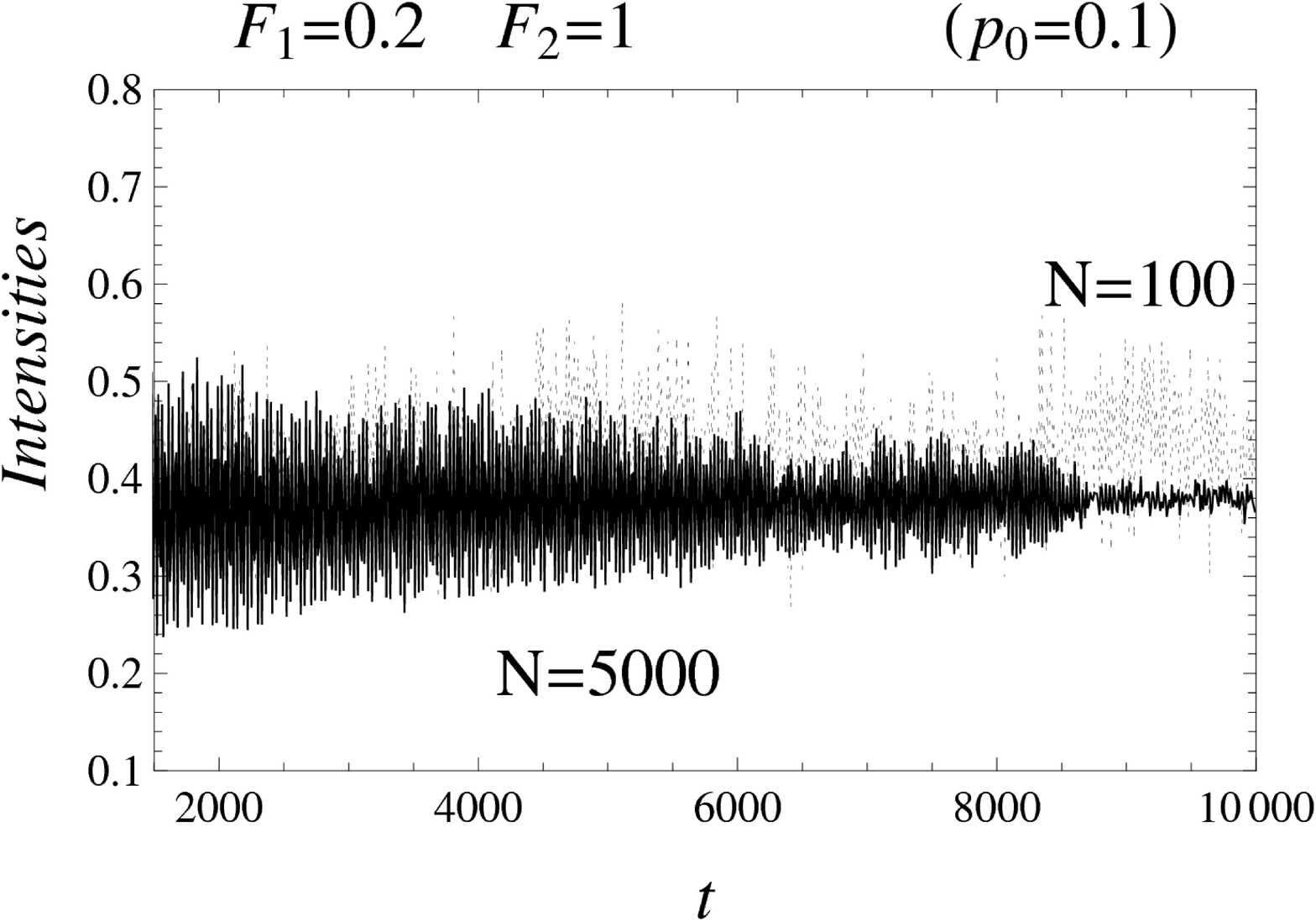}
\caption{Plots of the Langmuir-wave intensity $I_2$ for a long time evolution with $N=100$ (solid) and $N=5000$ (dotted). The left panel represents a zoom of the right one where the intensity values are averaged over $\Delta t=10$.}
\label{QSS}
\end{figure}

As discussed above, after the exponential instability, the intensity of the leading harmonics oscillates around a well-defined mean value. At this point, it is important to prove that such oscillations correspond to the attainment of a QSS which typically sets in in presence of long-range interactions \cite{B04,PLR04,A07,C06}. \figref{QSS} suggests that, by increasing the number of particles, the system is kept out of thermal equilibrium in a QSS even if one could be able to numerically follow the time evolution for very long times. \emphr{The process is driven by finite-$N$ effects and the relaxation increases when the system size $N$ is made large, and eventually diverges in the infinite-$N$ limit. Hence the relaxation to the statistical mechanics equilibrium is a consequence of the inherent granularity of the inspected medium.}
This issue is at the basis of the theoretical analysis of the next Section. The Lynden-Bell approach implies that this peculiar state corresponds to the statistical equilibrium of the continuum system derived from \erefs{GWM} and this statement will be corroborated in what follows. In fact, since the numerical simulations clearly show the existence of the QSSs, the mean values of the parameters derived by a specific numerical simulation are to be matched with those  ones computed by means of the theoretical analysis of the statistical equilibrium in the $N\to\infty$ limit.

\section{The Lynden-Bell approach and the Vlasov equation: the QSS distribution function}
\label{Vlasov-LyndenBell}

We now discuss a theoretical approach aimed at interpreting the appearance of the QSS phenomenology. To begin with, let us recall that the so-called Braun and Hepp Theorem \cite{BH77} states that, for a wide class of (long-range) interaction potentials, the $N$-body solution converges to the solution of the associated Vlasov equation in the limit $N\to\infty$ and at fixed volume. 
In this limit it is not necessary to follow all the particle trajectories because the evolution of the system  can be completely determined through the distribution function $f(p,x,t)$. In other words, the Vlasov equation provides the correct interpretative framework for $N\to\infty$. Within this picture, we analyze here the Lynden-Bell ansatz \cite{LB67} surmising that the QSSs correspond to the statistical equilibrium of the Vlasov equation. Thus, a QSS is the most probable macro-state which corresponds to the largest number of microscopic configurations thus maximizing the entropy. The main issue of this approach is that of providing a partition function
to describe the QSS.

In the following, we adapt the statistical model given in \cite{CNSNS} to the case under consideration. The Vlasov-Poisson system is rigorously obtained by taking the continuum limit of Eqs.\reff{EqMotionSim}, what gives
\begin{subequations}\label{VlasovEq}
\begin{align}
&\frac{\p f}{\p t}=-p\frac{\p f}{\p x}
+2\sum_{n} F_n [\Ex\cos(k_{n}x)-\Ey\sin(k_{n}x)]\,\frac{\p f}{\p p}\;,\label{eq:VlasovA}\\
&\frac{d \Ex}{d t}= F_n \iint f\cos(k_{n}x)\;dxdp\;,\phantom{\sum_{n} F_n}\label{eq:VlasovB}\\
&\frac{d \Ey}{d t}= -F_n \iint f\sin(k_{n}x)\;dxdp\;,\label{eq:VlasovC}
\end{align}
\end{subequations}
where the following quantities are conserved
\begin{subequations}\label{constraints}
\begin{align}
&\epsilon(f,E)=\iint\frac{p^2}{2}\,f\;dxdp+
2\sum_{n}\frac{F_n}{k_n}\iint f\,[\Ex\sin(k_{n}x)+\Ey\cos(k_{n}x)]\;dxdp\;,\label{vlasovenergy}\\
&\sigma(f,E)=\iint pf\;dxdp+\sum_{n} F_n [\Ex\,^{2}+\Ey\,^{2}]\;,\label{vlasovmomentum}\\
&m=\iint f\;dxdp\equiv 1\;,\phantom{\sum_{n} F_n}\label{vlasovmass}
\end{align}
\end{subequations}
which correspond to the pseudo-energy per particle, the momentum per particle and the mass, respectively.

As already shown, in the continuum limit ($N\to\infty$) the system remains trapped in a QSS, whence Lynden-Bell assumes \cite{LB67} that the QSSs can describe the equilibrium states of the Vlasov equation \reff{eq:VlasovA} and the statistical mechanics approach is used to predict the values of the macroscopic observables of the system at saturation. The subtle point of this scheme consists of making the difference between the two non-commuting ways of performing the two limits of  $N\to\infty$ and $t\to\infty$. Different asymptotic dynamical regimes stem from performing first the limit for large time and ``then'' for a large particle number and, on the other hand, performing first the limit for large $N$ before $t\to\infty$. In the former case the thermodynamical equilibrium is safely reached, while in the latter case the system is indefinitely trapped in a QSS, which is the regime that  we want to consider here.

Another crucial aspect of this theoretical analysis relies on the fact that the support of the distribution function $f(x,p,t)$ becomes progressively more filamented and stirred at very small scales evolving with \erefs{VlasovEq}, without attaining any final stable equilibrium. In this respect, the model involves a coarse-grained function $\fb$, obtained by averaging $f$ over finite size macrocells and a sudden stabilization is observed to follow the initial mixing phase of ``violent relaxation'' \cite{LB67}. Accordingly with the numerical simulations, we consider as initial conditions a water-bag distribution, which correspond to a uniformly occupied rectangular region in the phase space $(x,p)$, \ie
\begin{align}\label{WBform}
f(x,p,0)=
\left\{
\begin{array}{ll}
\fo=1/4x_0 p_0\;\quad & \textrm{if}\;-p_0<p<p_0\;\;\textrm{and}\;-x_0<x<x_0\;,\\
0\; \quad & \textrm{otherwise}\;.
\end{array}
\right.
\end{align}
\emphr{This simplified class of initial conditions, \ie the single-level water-bag, could be in principle extended to other more general types and, in this sense, a first develop can be found in \cite{AF12}, where the adequacy of the Lynden-Bell theory has been successfully tested for a multi level water-bag initial distribution.} 

Following the standard principles of statistical mechanics, we search for an  equilibrium state of the system as the state that corresponds to the most probable macrostate, \ie the one that maximizes the mixing entropy also satisfying all the constraints imposed by the dynamics. The mixing entropy is defined as the logarithm of the number $W$ of microscopic configurations corresponding to the same macroscopic state characterized by the probability density $\fb(x,p)$ \cite{LB67,CSR96}. It reads
\begin{align}
s=\ln [W({\fb})]\;.	
\end{align}
In order to evaluate this quantity, the basic idea is to discretize the (bounded) phase space into a finite number of macrocells and then to divide such macrocells into microcells of finite size. Moreover, since in the system \reff{VlasovEq} all the general quantities defined by $C_h=\int f^h dxdp$ are conserved, \emphr{the fine-grained distribution $f$ remains two-level. Thus, setting the size of the microcells at a scale smaller than the fine-grained distribution, these can be either occupied by the level $\fo$ or $0$, one excluding the other, similarly to a Pauli-like principle.} After combinatorial analysis, the entropy $s$ can be written as \cite{CSR96}
\begin{align}\label{vlasoventropy}
s(\bar{f})=-\int\Big[\frac{\fb}{\fo}\;\ln\Big(\frac{\fb}{\fo}\Big)
+\Big(1-\frac{\fb}{\fo}\Big)\;\ln\Big(1-\frac{\fb}{\fo}\Big)\Big]\;dxdp\;.
\end{align}
The analogy with the Fermi-Dirac statistics follows from the presence of the second term in the expression above that accounts for the exclusion principle (the similarity is indeed purely formal since the contexts are strongly different). Furthermore, it is worth underlying that, in the dilute limit for $\fo\to\infty$, this additional term can be neglected recovering the usual Boltzmann-Gibbs statistics. \emphr{An interesting discussion on the statistical mechanics approach to Vlasov systems is given in \cite{D82}, where the Lynden-Bell strategy is evoked among different possibilities. Such an analysis is however carried out assuming the Maxwell-Boltzmann entropy, which holds in the diluted limit.} 

Since we aim to compare the numerical results with the Lynden-Bell analysis, we consider the evolution of the system in the presence of only two harmonics $n=1,2$. The equilibrium distribution is derived by maximizing the expression \reff{vlasoventropy} and imposing the dynamical constraints. This corresponds to solving the following variational problem:
\begin{equation}\label{valsoventropy2}
S(p_0)=\max_{\fb,\Ex,\Ey}\;\;s(\fb)\Bigg|
_{\epsilon(\fb,\Ex,\Ey)=\epsilon_0=p_0^{2}/6;\;\;\sigma(\fb,\Ex,\Ey)=p_0;\;\;m(\fb,\Ex,\Ey)=1}\;\;,
\end{equation}
which leads to the equilibrium values
\begin{subequations}\label{equilibriumvalues}
\begin{align}
\fqss &=\fo \frac{e^{[-\beta(p^2/2+2\sum_{n} F_n[\Ey\cos(k_{n}x)+\Ex\sin(k_{n}x)]/k_n)-\lambda p -\mu]}}
{1+e^{[-\beta(p^2/2+2\sum_{n} F_n[\Ey\cos(k_{n}x)+\Ex\sin(k_{n}x)]/k_n)-\lambda p -\mu]}}\;,\phantom{\sum_{n} F_n}\label{fBar}\\
\Ex&=-\frac{\beta F_n}{k_n \lambda} \iint \fb \sin(k_{n}x)\;dxdp\;,\phantom{\sum_{n} F_n}\\
\Ey&=-\frac{\beta F_n}{k_n \lambda} \iint \fb \cos(k_{n}x)\;dxdp\;,
\end{align}
\end{subequations}
where $\beta/f_0$, $\lambda/f_0$ and $\mu/f_0$ are the Lagrange multipliers for the energy, momentum and normalization (mass) constraints, respectively. Substituting the equations above into \erefs{constraints}, one obtains the following system of self-consistent equations
\begin{subequations}\label{VlasovSys}
\begin{align}
&\lambda=\beta[(E_1^x)^{2}+(E_2^x)^{2}+(E_3^y)^{2}]\;,\phantom{\int}\\
&\frac{\fo\gamma}{\sqrt{\beta}}\int dx\;e^{z}M_1(\gamma e^{z})=1\;,\\
&\frac{\fo\gamma}{2\sqrt{\beta^{3}}}\int dx\;e^{z}M_2(\gamma e^{z})=
p_0^{2}/6+\frac{3}{2}\,(|E_1|^{2}+|E_2|^{2})^{2}\;,\\
&\frac{\fo\gamma}{\sqrt{\beta}}\int dx\;\sin x\,e^{z}M_1(\gamma e^{z})=
-(|E_1|^{2}+|E_2|^{2})\;E_1^{x}/F_1\;,\\
&\frac{\fo\gamma}{\sqrt{\beta}}\int dx\;\sin(2x)\,e^{z}M_1(\gamma e^{z})=
-2(|E_1|^{2}+|E_2|^{2})\;E_2^{x}/F_2\;,\\
&\frac{\fo\gamma}{\sqrt{\beta}}\int dx\;\cos(2x)\,e^{z}M_1(\gamma e^{z})=
-2(|E_1|^{2}+|E_2|^{2})\;E_2^{y}/F_2\;,
\end{align}
\end{subequations}
where, since the system is invariant under a phase translation $x\to x+\phi$, we have set, without loss of generality, $\int\fb\cos x=0$ which implies $E_1^y=0$, and where we have defined
\begin{subequations}
\begin{align}
&\gamma=e^{\lambda^{2}/2\beta-\mu}\;,
\phantom{\int du\;\frac{u^{2}e^{-u^{2}/2}}{1+\gamma e^{z}e^{-u^{2}/2}}}\\
&z=-\beta[2F_1 E_1^x \sin x+
F_2(E_2^y\cos(2x)+E_2^x\sin(2x))]\;,
\phantom{\int du\;\frac{u^{2}e^{-u^{2}/2}}{1+\gamma e^{z}e^{-u^{2}/2}}}
\\
&M_1(\gamma e^{z})=\int du\;\frac{e^{-u^{2}/2}}{1+\gamma e^{z}\epsilon-u^{2}/2}\;,\\
&M_2(\gamma e^{z})=\int du\;\frac{u^{2}e^{-u^{2}/2}}{1+\gamma e^{z}e^{-u^{2}/2}}\;.
\end{align}
\end{subequations}

The continuum dynamics is than described by the system above that can be solved numerically, using a Newton-Raphson method, to determine the values of the amplitudes of the two harmonics and the Lagrange multipliers.
\begin{figure}[!ht]
\centering
\includegraphics[width=0.45\hsize]{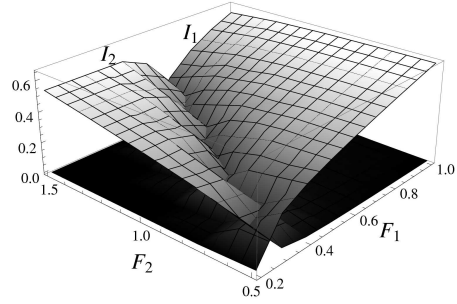}
\caption{Integration of the system \reff{VlasovSys} for different values of $F_n$.}
\label{In3D}
\end{figure}
This algorithm solves the system for the six variables $\beta,\;\lambda,\;\mu,\;E_1^x,\;E_2^x,\;E_2^y$, and the integrals contained in the equations are performed numerically. In this picture, once having fixed the initial conditions and the free parameters (the coupling constants), the numerical integration provides a single value for the mode intensity $I_n=\modu{E_n}^{2}$ which has to be compared to the averaged value of the simulation results and this is the aim of the next Section.

Let us now limit the $F_n$-parameter space to a finite portion of the plane $[0,1]\times[0,3]$. As is done in the numerical simulations, we consider a homogeneous initial condition, as specified by \ereff{WBform}, with $x_0=\pi$ and $p_0=0.1$. Within the explored region in the plane $(F_1,F_2)$, the Lynden-Bell theory provides two distinct stationary equilibria, one with $E_1\neq0$, $E_2\simeq0$ (where $I_1$  dominates and $I_2$ is negligible), and the other one with $E_1\simeq0$, $E_2\neq0$ (only $I_2$ is amplified). These two dynamical regimes are associated to two distinct regions, as can be seen in  \figref{In3D}, where these regions are divided by a separatrix which corresponds to the intersection line of the two planes representing the mode intensity amplitudes. The system undergoes a first-order phase transition when it passes from a zone to the other one.
\begin{figure}[!ht]
\centering
\includegraphics[width=0.49\hsize]{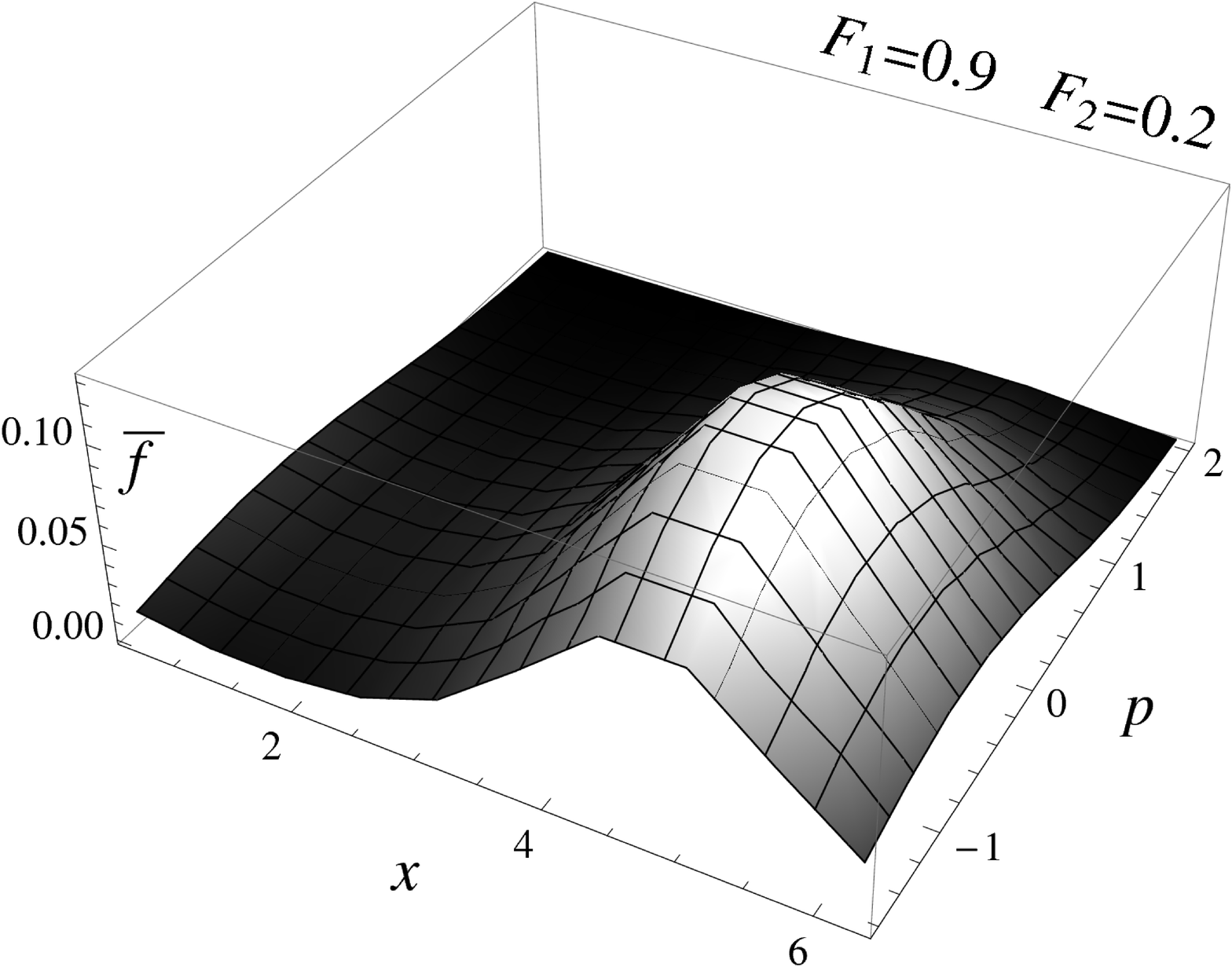}
\includegraphics[width=0.49\hsize]{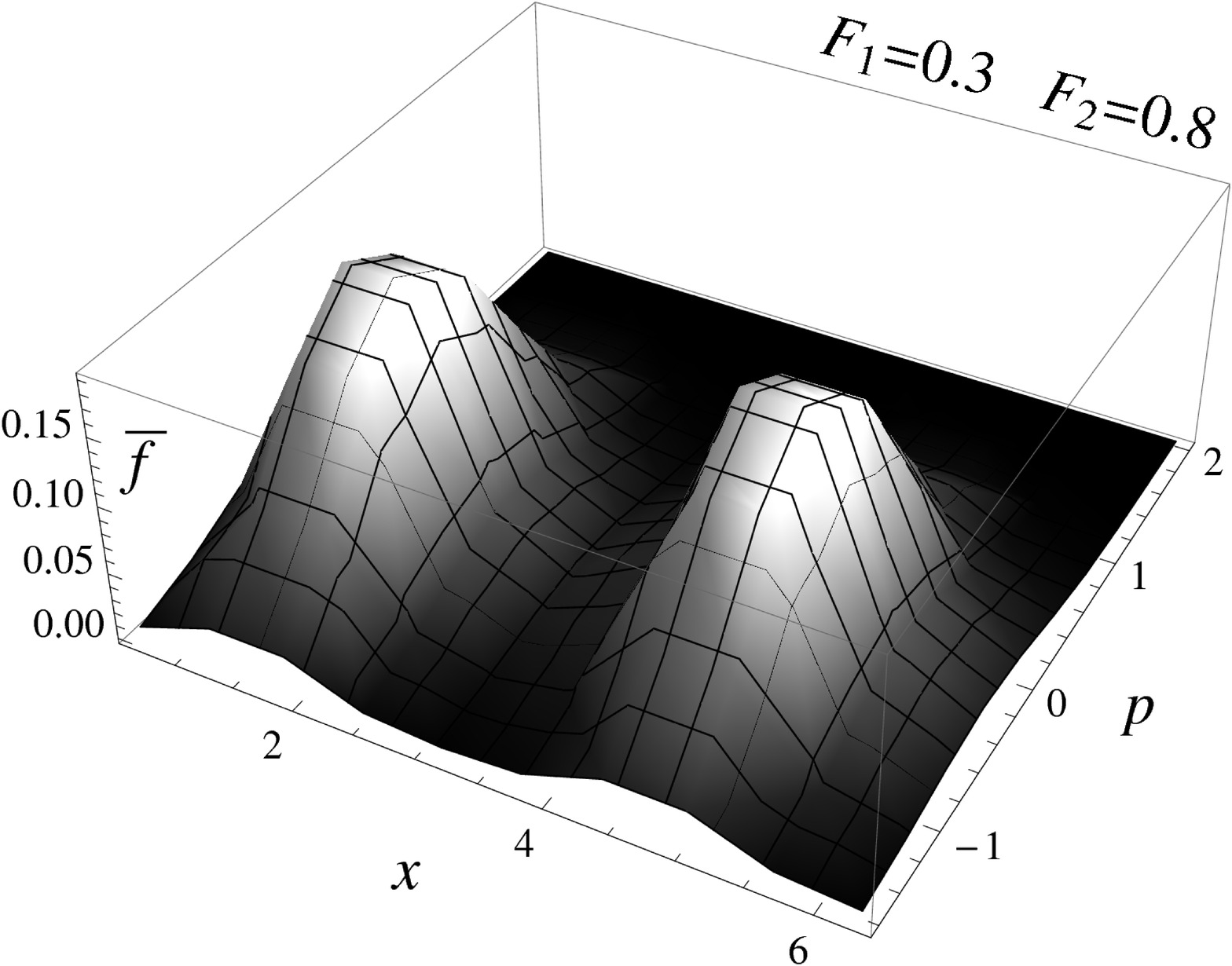}
\caption{Distribution function \ereff{fBar} for two different pairs of $F_n$.}
\label{fbqss}
\end{figure}

The evaluation of $E_n$ allows for the construction of the distribution function \ereff{fBar} (see \figref{fbqss}) which has been derived from the equilibrium principle and, as the theoretical prediction on the observables of the model is confirmed by the simulations, it can consistently characterize the QSSs features. Comparing the distribution function profile with the corresponding predictions of the discrete system (lower panels of \figref{water-bag-evolution}), a clear correspondence emerges accordingly to the Lynden-Bell conjecture.

\begin{figure}[!ht]
\centering
\includegraphics[width=0.39\hsize]{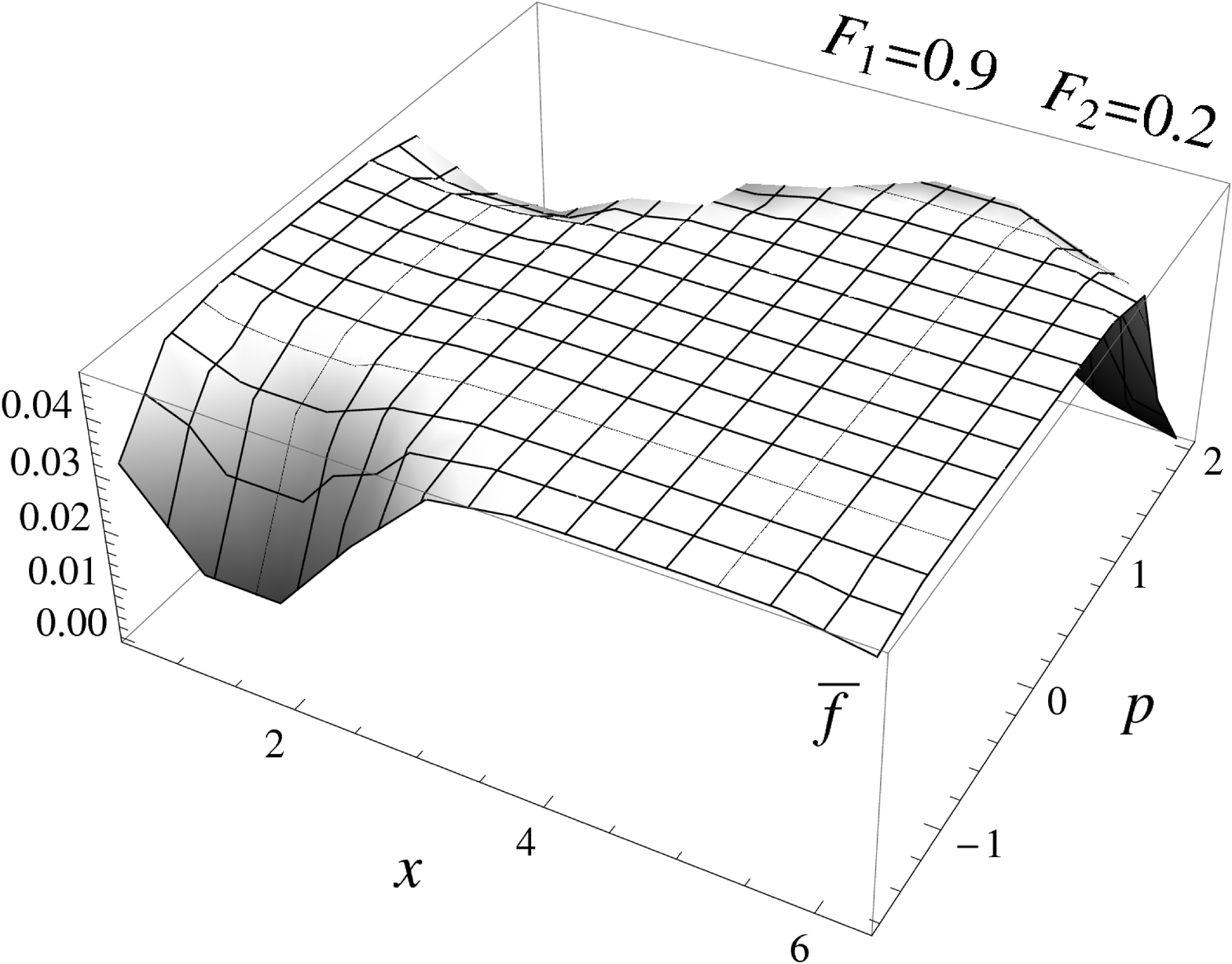}
\caption{Distribution function \ereff{fBar} for $p_0=1.3$.}
\label{fR}
\end{figure}

As discussed in Section \ref{RES}, increasing the value of $p_{0}$, the thermalization of the beam is weaker and, if the critical threshold is exceeded, the whole system does not respond since the particles are not trapped by the wave. The distribution function for an above-threshold initial beam-particles momentum spread $p_0=1.3$, as in the left panel of \figref{NoRes}, is reported in \figref{fR} and it is seen to be consistent with the previous analysis.

\emphr{We conclude this Section by discussing the role of the initial conditions assumed in the present treatment. It is important to recall that the Lynden-Bell analysis that we have developed applies to a specific class of the initial value problem: the so called water-bag. Particles are assumed to be uniformly distributed inside a compact domain of phase space, not necessarily a rectangle. This has to be regarded as a simplifying assumption, which allows for straightforward analytical manipulation and, when it comes to the comparison with direct experiments, the initial condition assumed is too schematic.
In this respect, we remark that, usually, one can build the a distribution function performing a variable transformation in the action/angle space and the result is the bump on tail distribution, \ie a Maxwellian plus a ``beam (smooth) function''. Another common approach consists in replacing such distribution by a linear function in the velocity space, centered around a finite velocity $v^{*}$. In this scheme, the resonance occurs (if the mode frequency is not small) where the condition $\omega_n=k_nv^{*}$ can be satisfied (we recall that $\omega_n$ and $k_n$ are the $n$-mode pulsation and wavenumber, respectively) and where the slope of the distribution function results to be positive. In our work, the mode frequencies are assumed to be small, thus the linear Landau resonance occurs at $v=p=0$, in agreement whit the chosen domain. This is equivalent to shift the velocity as $v=v-v^{*}$. The treated distribution function is not linear but rather a step function and the resulting destabilizing mechanism is quite far from the standard (linear) picture for the bump on tail instability.}

\emphr{However, the general method that we have developed, and successfully tested versus numerical simulations, can be extended to describe more complex, and possibly more realistic, scenarios. For instance, as discussed above, any given initial velocity distribution could be adequately approximated via a step-wise function that involves an arbitrarily large set of different levels \cite{AF12}. The Lynden-Bell analysis can be readily adapted to scrutinize such a general scenario, at the price of an augmented computational effort.}

\section{Test of the Lynden-Bell approach}
\label{Test}

\begin{figure}[!ht]
\centering
\includegraphics[width=0.49\hsize]{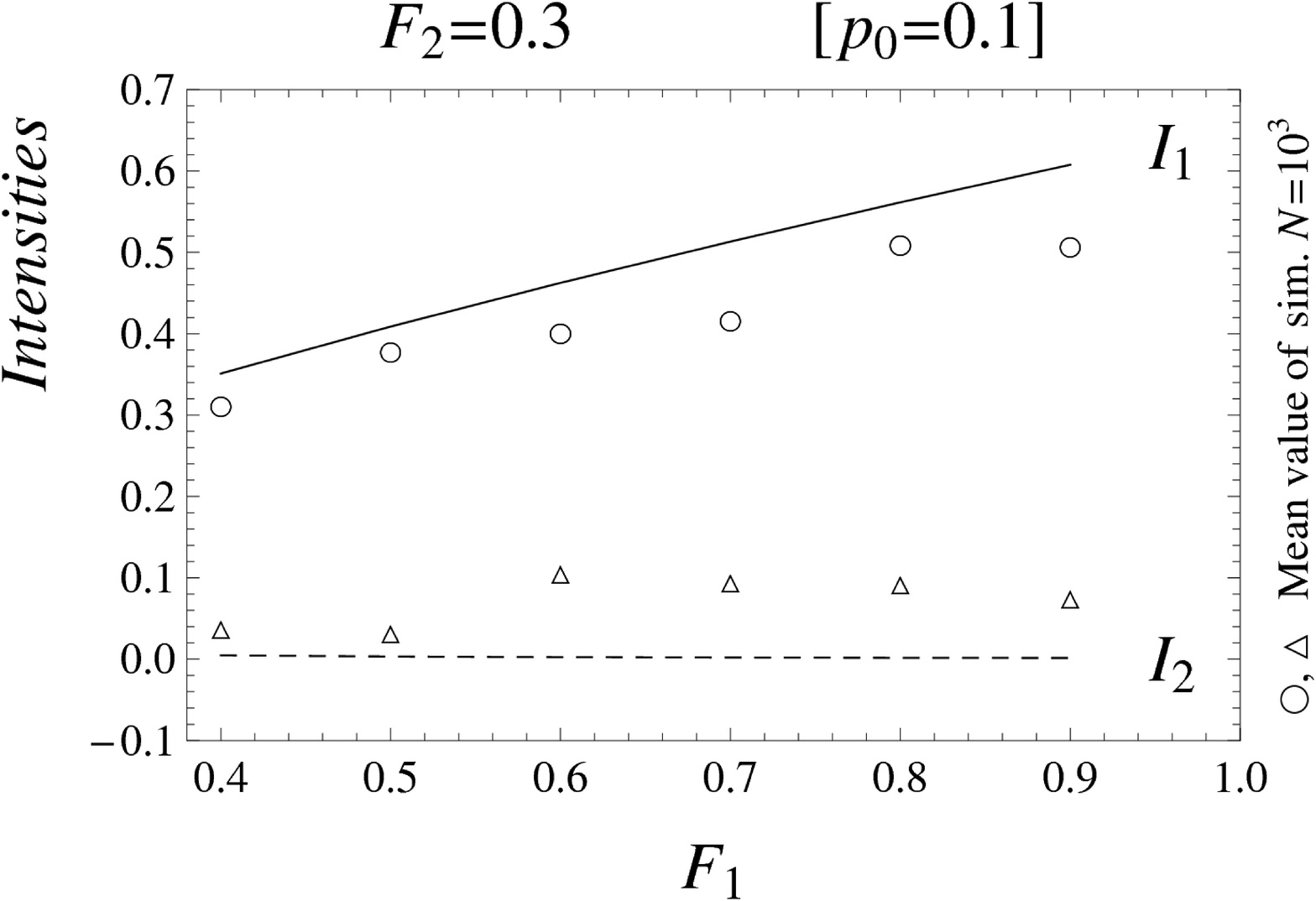}
\includegraphics[width=0.49\hsize]{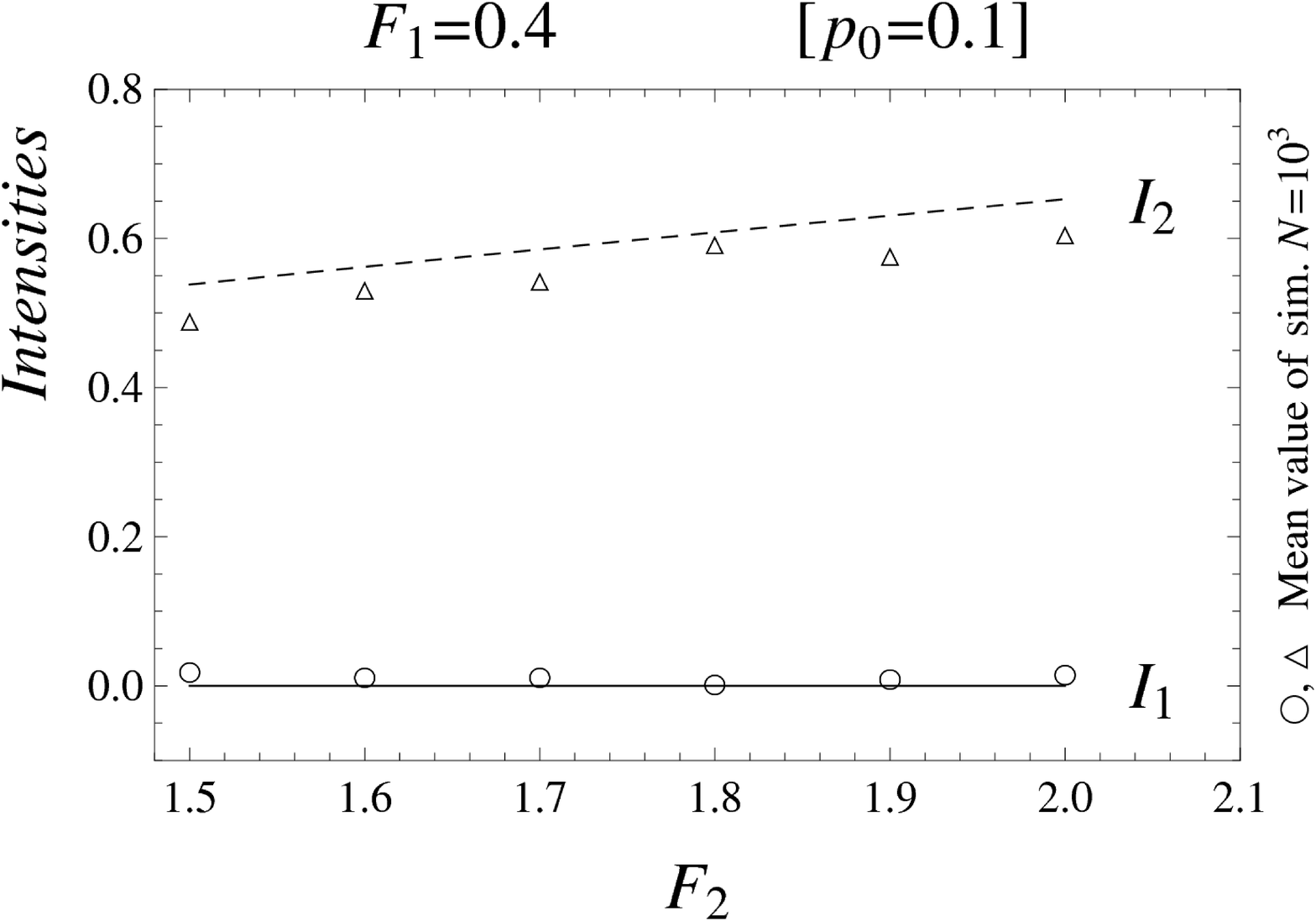}
\caption{Test of the Lynden-Bell predictions for the wave intensities. The lines (solid and dashed) indicate the integration of the Lynden-Bell theoretical system \reff{VlasovSys} for different values of $F_n$; while each symbol represents the mean value of $I_n$ (averaged for $25<t<80$) calculated by a single simulation with $N=10^{3}$.}
\label{LBtest}
\end{figure}

In order to test the theoretical predictions of the previous Section, we now treat two different sets of the parameter values which give rise to distinct scenarios. In particular, we fix one coupling constant and we vary the other one in order to explore different values of the mode intensities. The integration of the continuum system is performed for a large number of values of the running coupling constant and the results are represented with a solid line in \figref{LBtest}.
\emphr{At variance, the numerical simulations of the discrete system ($N=10^{3}$) are run for a limited number of distinct values of the $F_n$ and until a time $t=80$. At this time, the coherent clumps are well-formed (as represented in \figref{water-bag-evolution}) providing a sufficient sampling of the dynamics. In fact, the intensity oscillations (plotted in \figref{intensities}) are stabilized and the corresponding averaged value can reasonably represent the effective long-time evolution. Moreover, \figref{QSS} shows how a beam composed by a number of particles of about $10^{3}$ remains effectively trapped in the out-of-equilibrium state (a QSS) at least until a time of order $t\sim8\times10^{3}$. After that time, the system progressively evolve toward the thermodynamical equilibrium. In this respect, the simulations samplings can be considered undoubtedly in a QSS.}

As a result, the numerical analysis confirms the theoretical scenario: the simulations show the emergence of the two regions in the parameter space, where the average intensities match the Lynden-Bell predictions. In \figref{LBtest}, it is shown how each simulated mean value of the intensity $I_n$ (averaged from $25<t<80$ in order to neglect the first phase of violent-relaxation evolution) well overlaps the intensity calculated by the numerical integration of the system \reff{VlasovSys}.

Concluding, the test of the Lynden-Bell ansatz here proposed demonstrates how the QSSs, predicted in long-range interactions and numerically simulated with the system \reff{EqMotionSim}, are consistently characterized by the Lynden-Bell-Vlasov expressions \reff{equilibriumvalues}. In other words, one can state that \ereff{fBar} (plotted in \figref{fbqss}) effectively represents the distribution function of the QSSs. In this sense, we want to stress that, using the tested Lynden-Bell approach, it results possible to consistently characterize the non-equilibrium microphysics of the system through a statistical equilibrium analysis.

\section{Conclusions}
\label{Conclusions}

The beam-plasma instability issue has been tackled by means of numerical simulations of the discrete system and through a statistical analysis of the corresponding continuum Vlasov model. This specific problem is properly described by long-range interactions and it has been shown how the system lives in out-of-equilibrium states (QSSs) whose lifetimes diverge with the number of particles in the system. A theoretical characterization of these peculiar dynamical regimes has been addressed using the Lynden-Bell approach which surmises that the QSSs are the statistical equilibria of the Vlasov equation. The distribution function expected to characterize these states has been also derived. The main result of our work consists of the numerical proof that the Lynden-Bell distribution function correctly predicts the amplitudes of the electrostatic modes and thus effectively describes the QSSs. This verification allows to consistently derive the macroscopical observables of the system despite its non-linear out-of-equilibrium nature.

We have also phenomenologically studied the effective thermalization of the system and the simulations have shown the existence of a threshold value of the initial spread of beam-particle momenta which sharply discriminates between the resonant and non-resonant regimes. Moreover, by considering an initial localization of particles in the resonant portion of the phase space, we have outlined how a very small increase of the number of the fully resonant particles is sufficient to get the resonant response of the system also in the case of out-of-resonance initial conditions.

\emphr{The simulations and the theoretical analysis have shown the presence of two distinct zones, in the parameter space, associated to a different leading mode, respectively. The system undergoes a first-order phase transition when passing from one zone to the other and peculiar features should emerge near the separatrix between the two dynamical regimes. In this respect, a common striking experimental observation results in the \emph{frequency chirping} phenomenon, \ie a change of the mode frequency with time \cite{BBP96,BBP97,BB99} (for a numerical test of the model, see \cite{VB07}, and a comparison to experimental data was performed in \cite{L10}).
}

\emphr{An important point regarding the discrete analysis should be emphasized: at variance with the simulations reported in this work, the Particle-in-cell (PIC) or Eulerian models are usually performed in the weak drive limit, with only one unstable mode. In this scheme, the linear (or non linear) growth rate is smaller than the real part of the pulsation, while we deal with the strong drive limit. From \figref{intensities}, it can be argued that the growth rate in the initial phase is bigger than the oscillation frequency (clump rotation). Also the model \cite{BBP96} does not treat modes and particles on equal footing}.

As a concluding remark, even though our present model is far from directly applying to the problem of the $\alpha$-particle thermalization in a burning plasma, it nevertheless points out that the dynamics of a population of fast particles interacting with a thermalized plasma can be definitely non trivial.

\ack{MP carried out the present work while on leave of absence from \emph{Osservatorio Astrofisico di Arcetri} - INAF, Firenze, Italy.}


\section*{References}
\begin{thebibliography}{99}

\bibitem{A09}
A. Campa, T. Dauxois, S. Ruffo,
\emph{Phys. Rept.} \textbf{480}, 57 (2009).

\bibitem{B04}
J. Barr\'e et al.,
\emph{Phys. Rev. E} \textbf{69}, 045501 (2004).

\bibitem{PLR04}
A. Pluchino, V. Latora, A. Rapisarda,
\emph{Phys. Rev. E} \textbf{69}, 056113 (2004).

\bibitem{SK96}
H. Schamel, J. Korn,
\emph{Physica Scr.} \textbf{T63}, 63 (1996).

\bibitem{LI12}
M. Lesur, Y. Idomura,
\emph{Nucl. Fus.} \textbf{52}, 094004 (2012).

\bibitem{LB67}
D. Lynden-Bell, \emph{Mon. Not. RAS} \textbf{136}, 101 (1967).

\bibitem{LPT08}
Y. Levin, R. Pakter, T.N. Teles,
\emph{Phys. Rev. Lett.} \textbf{100}, 040604 (2008).

\bibitem{OWM71}
T.M. O'Neil, J.H. Winfrey, J.H. Malmberg, 
\emph{Phys. Fluids} \textbf{14}, 1204 (1971).

\bibitem{EEb}
Y. Elskens, D.F. Escande,
\emph{Microscopic Dynamics of Plasmas and Chaos}, IoP Publishing, 2003.

\bibitem{BB99}
H.L. Berk et al.,
\emph{Phys. Plasmas} \textbf{6}, 3102 (1999).

\bibitem{TMM94}
J.L. Tennyson, J.D. Meiss, P.J. Morrison,
\emph{Physica D} \textbf{71}, 1 (1994).

\bibitem{AEE98}
M. Antoni, Y. Elskens, D.F. Escande,
\emph{Phys. Plasmas} \textbf{5}, 841 (1998).

\bibitem{O65}
T.M. O'Neil,
\emph{Phys. Fluids} \textbf{8}, 2255 (1965).

\bibitem{M72}
N.G. Matsiborko et al.,
\emph{Plasma Phys.} \textbf{14}, 591 (1972).

\bibitem{S63}
V.D. Shapiro,
\emph{Sov. JETP} \textbf{17}, 416 (1963).

\bibitem{FE98}
M.-C. Firpo, Y. Elskens,
\emph{J. Stat. Phys.} \textbf{93}, 193 (1998).

\bibitem{FE00}
M.-C. Firpo, Y. Elskens,
\emph{Phys. Rev. Lett.} \textbf{84}, 3318 (2000).

\bibitem{F01}
M.-C. Firpo et al.,
\emph{Phys. Rev. E} \textbf{64}, 026407 (2001).

\bibitem{OM68}
T.M. O'Neil, J.H. Malmberg,
\emph{Phys. Fluids} \textbf{11}, 1754 (1968).

\bibitem{MK78}
H.E. Mynick, A.N. Kaufman,
\emph{Phys. Fluids} \textbf{21}, 653 (1978).

\bibitem{AF06}
A. Antoniazzi, D. Fanelli, Y. Elskens, S. Ruffo,
\emph{Eur. Phys. J. B} \textbf{50}, 603 (2006).

\bibitem{A05}
A. Antoniazzi, G. De Ninno, D. Fanelli, A. Guarino, S. Ruffo,
\emph{J. Phys.: Conf. S.} \textbf{7}, 143 (2005).

\bibitem{FLA06}
M.-C. Firpo, F. Leyvraz, G. Attuel,
\emph{Phys. Plasmas} \textbf{13}, 122302 (2006).

\bibitem{CNSNS}
A. Antoniazzi, R.S. Johal, D. Fanelli, S. Ruffo,
\emph{Comm. Nonlin. Sci. Num. Sim.} \textbf{13}, 2 (2008).

\bibitem{DIIb}
P.H. Diamond, S.-I. Itoh, K. Itoh,
\emph{Modern Plasma Physics -- Volume 1: Physical Kinetics of Turbulent Plasmas},
Cambridge University Press (2010).

\bibitem{BNR70}
H.L. Berk, C.E. Nielsen, K.V. Roberts,
\emph{Phys. Fluids} \textbf{13}, 980 (1970).

\bibitem{Neg05}
D. del-Castillo-Negrete,
\emph{Plasma Phys. Control. Fus.} \textbf{47}, A53 (2005).

\bibitem{LP81}
I.M. Lifshitz, L.P. Pitaevski, 
\emph{Physical Kinetics}, Pergamon Press (1981).

\bibitem{B84}
R. Bonifacio et al.,
\emph{Opt. Commun.} \textbf{50}, 373 (1984).

\bibitem{A07}
A. Antoniazzi et al.,
\emph{Phys. Rev. E} \textbf{75}, 011112 (2007).

\bibitem{C06}
P.H. Chavanis,
\emph{Euro Phys. J. B} \textbf{53}, 487 (2006).

\bibitem{BH77}
W. Braun, K. Hepp,
\emph{Comm. Math. Phys.} \textbf{56}, 101 (1977).

\bibitem{AF12}
M. Assllani et al.,
\emph{Phys. Rev. E} \textbf{85}, 021148 (2012).

\bibitem{CSR96}
P.H. Chavanis, J. Sommeria, R. Robert,
\emph{ApJ} \textbf{471}, 385 (1996).

\bibitem{D82}
T.H. Dupree,
\emph{Phys. Fluids} \textbf{25}, 277 (1982).

\bibitem{BBP96}
H.L. Berk, B.N. Breizman, M. Pekker,
\emph{Phys. Rev. Lett.} \textbf{76}, 1256 (1996).

\bibitem{BBP97}
H.L. Berk, B.N. Breizman, N.V. Petiashvili,
\emph{Phys. Lett. A} \textbf{234}, 213 (1997).

\bibitem{VB07}
R.G.L. Vann, H.L. Berk, A.R. Soto-Chavez,
\emph{Phys. Rev. Lett.} \textbf{99}, 025003 (2007).

\bibitem{L10}
M. Lesur et al.,
\emph{Phys. Plasmas} \textbf{17}, 122311 (2010).





\end{thebibliography}
\end{document}